\documentclass{aa}
\usepackage{graphicx}       
\usepackage{natbib}

\def\asca{{\it ASCA~\/}}

\def\chandra{{\it Chandra~\/}}

\def\Msun{\hbox{$\rm ~M_{\odot}$}}
\def\mmdo{{M$_{\rm MDO}$}}

\def\H0{{\rm ~km~s^{-1}~Mpc^{-1}}}

\def\hii{\ion{H}{II}}
\def\ha{{H$\alpha$}}
\def\hanii{H$\alpha$+[N\,{\small~II}]}
\def\fullhanii{H$\alpha$+[N\,{\small~II}]~$\lambda\lambda$6548,6583}
\def\oi{[\ion{O}{I}]}
\def\oiii{[\ion{O}{III}]}
\def\nii{[\ion{N}{II}]}
\def\sii{[\ion{S}{II}]}
\def\fulloi{[\ion{O}{I}]$\,\lambda$6300}
\def\fulloiii{[\ion{O}{III}]$\,\lambda$5007}
\def\fullnii{[\ion{N}{II}]$\,\lambda$6583}

\def\Loi{L$_{\rm [O\,{I}]}$}       
\def\Lha{L$_{\rm H\alpha}$}       
\def\Lhanii{L$_{\rm H\alpha+[N\,{II}]}$}

\def\.25{0.25 keV\thinspace}

\def\d19{D$\,\leq\,$19~Mpc} 
\newcommand{\sgra}{Sgr~A*}

\let\lesssim=\la
\let\gtrsim=\ga

\newcommand\dummytable{\refstepcounter{table}}%

\begin{document}

\title{Radio Sources in Low-Luminosity Active Galactic Nuclei.}
\subtitle{III. ``AGNs'' in a Distance-Limited Sample of ``LLAGNs''}

\author{
   Neil M. Nagar
   \inst{1}
   \and
   Heino Falcke
   \inst{2}
   \and 
   Andrew S. Wilson
   \inst{3}
   \and
   James S. Ulvestad
   \inst{4}
}
\offprints{Neil M. Nagar}
\institute{Arcetri Observatory, Largo E. Fermi 5,
             Florence 50125, Italy \\
             \email{neil@arcetri.astro.it}
           \and
           Max-Planck-Institut f\"{u}r Radioastronomie,
             Auf dem H\"{u}gel 69,
	     53121 Bonn, Germany \\
             \email{hfalcke@mpifr-bonn.mpg.de}
           \and
           Department of Astronomy, University of Maryland,
             College Park, MD 20742, U.S.A. \\
           Adjunct Astronomer, Space Telescope Science Institute,
             3700 San Martin Drive, Baltimore, MD 21218, U.S.A. \\
             \email{wilson@astro.umd.edu} 
           \and
           National Radio Astronomy Observatory, P.O. Box 0, 
	     Socorro, NM 87801, U.S.A. \\
             \email{julvesta@nrao.edu}
          }

\date{Received January 23, 2002; accepted June 7, 2002}

\abstract{
This paper presents the results of a high resolution radio imaging survey
  of all known (96) low-luminosity active galactic nuclei (LLAGNs) at \d19. 
  We first report new 2~cm (150~mas resolution using the VLA) and 6~cm 
  (2~mas resolution using the VLBA) 
  radio observations of the previously unobserved nuclei in our samples and 
  then present results on the complete survey.
We find that almost half of all LINERs and low-luminosity Seyferts have
  flat-spectrum radio cores when observed at 150~mas resolution. 
Higher (2~mas) resolution observations of a flux-limited subsample have 
  provided a 100\% (16 of 16) detection rate of pc-scale radio cores, with 
  implied brightness temperatures $\gtrsim\,10^8\,$K. The five LLAGNs with the 
  highest core radio fluxes also have pc-scale `jets.'
Compact radio cores are almost exclusively found in massive ellipticals
  and in type~1 nuclei (i.e. nuclei with broad \ha\ emission). 
Only a few `transition' nuclei have compact radio cores; those detected
  in the radio have optical emission-line diagnostic ratios close to those of 
  LINERs/Seyferts. This indicates that some transition nuclei are truly 
  composite Seyfert/LINER+\hii\ region nuclei, with the radio core power 
  depending on the strength of the former component.
The core radio power is correlated with the nuclear optical `broad'
  \ha\ luminosity, the nuclear optical `narrow' emission-line 
  luminosity and width, and with the galaxy luminosity. 
  In these correlations LLAGNs fall close to the low-luminosity 
  extrapolations of more powerful AGNs.
  The scalings suggest that many of the radio-non-detected LLAGNs are simply
  lower power versions of the radio-detected LLAGNs.
The ratio of core radio power to nuclear optical emission-line luminosity 
  increases with increasing bulge luminosity for all LLAGNs. Also, there is evidence 
   that the luminosity of the disk component of the galaxy is correlated with the 
   nuclear emission-line luminosity (but not the core radio power). 
About half of all LLAGNs with multiple epoch data show significant inter-year 
   radio variability.  
Investigation of a sample of $\sim\,$150 nearby bright galaxies, most of them 
   LLAGNs, shows that the nuclear ($\leq$150~mas size) radio power is strongly 
   correlated with both the black hole mass and the galaxy bulge luminosity; 
   linear regression fits to all $\sim\,$150 galaxies give:  
   log P$_{\rm 2cm}$ = 1.31($\pm$0.16)~log~\mmdo\              $+\,8.77$ and
   log P$_{\rm 2cm}\,=$ 1.89($\pm$0.21)~log~L$_{\rm B}$(bulge) $-\,0.17$. 
Low accretion rates ($\leq10^{-2}-10^{-3}$ of the Eddington rate) are implied 
   in both advection- and jet-type models. 
In brief, all evidence points towards the presence of accreting massive black 
  holes in a large fraction, perhaps all, of LLAGNs, with the nuclear radio 
  emission originating in either the accretion inflow onto the massive black hole 
  or from jets launched by this black hole - accretion disk system. 
\keywords{accretion, accretion disks --- galaxies: active --- galaxies: jets
--- galaxies: nuclei --- radio continuum: galaxies --- surveys}
} 

\titlerunning{AGNs in a Distance-Limited Sample of LLAGNs}
\authorrunning{Nagar, N. M. et al.}

\maketitle

\section{Introduction}

The debate on the power source of low-luminosity active galactic 
nuclei (LLAGNs, i.e. low-luminosity Seyferts, LINERs, and ``transition'' 
nuclei [nuclei with spectra intermediate between those of LINERs and 
\hii\ regions]) is a continuing one. Their low emission-line luminosities 
\citep[L$_{\rm H\alpha}~\leq$ 10$^{40}$ erg s$^{-1}$ by definition;][hereafter H97a]{hoet97a}
can be modeled in terms of photoionization by hot, young stars 
\citep{termel85,filter92,shi92}, 
by collisional ionization in shocks \citep{kosost76,foset78,hec80,dopsut95}
or by aging starbursts \citep{aloet99}.

On the other hand, evidence has been accumulating that at least some 
fraction of LLAGNs share characteristics in common with their more powerful
counterparts - radio galaxies and powerful Seyfert galaxies.
These similarities include the presence of compact nuclear radio cores
\citep{hec80}, water vapor megamasers \citep{braet97},
nuclear point-like UV sources \citep{maoet95,baret98},
broad \ha\ lines \citep[][hereafter H97c]{hoet97c}, and broader
\ha\ lines in polarized emission than in total emission \citep{baret99}.
If LLAGNs are truly scaled down AGNs then the challenge is to explain their
much lower accretion luminosities. This requires either very low accretion 
rates ($\sim\,10^{-8}$ of the Eddington accretion rate, L$_{\rm Edd}$)
or radiative efficiencies (the ratio of radiated energy to accreted mass) 
much lower than the typical value of $\sim\,$10\% 
\citep[e.g.~Chapter~7.8 of][]{fraet95} assumed for powerful AGNs.    

One well-known property of some powerful AGNs is a compact 
(sub-parsec), flat-spectrum (usually defined as 
$\alpha~\geq$ $-$0.5, S$_\nu~\propto \nu^\alpha$)
nuclear radio source, usually interpreted as
the synchrotron self-absorbed base of the jet which fuels
larger-scale radio emission. Astrophysical jets are known to
be produced in systems undergoing accretion onto a compact
object \citep[e.g.][]{pri93,bla93} so such compact radio 
sources in galactic nuclei may reasonably be considered a 
signature of an AGN. Much theoretical work
\citep[e.g.][]{beget84,lovrom96,falbie99} has been devoted to 
this disk-jet relationship in the case of galactic nuclei
and it has been suggested that scaled-down versions of AGN 
jets can produce flat-spectrum radio cores in LLAGNs \citep{falbie99}.
Compact nuclear radio emission with a flat to inverted spectrum
is also expected from the accretion inflow in advection-dominated 
\citep[ADAF;][]{naret98} or convection-dominated \citep[CDAF;][]{naret00} 
accretion flows, possible forms of accretion onto a black hole at low accretion 
rates \citep{reeet82}. 
Flat-spectrum radio sources can also result through thermal emission
from ionized gas in normal \hii\ regions or through free-free
absorption of non-thermal radio emission, a process which probably
occurs in compact nuclear starbursts \citep{conet91}.
The brightness temperature, T$_{\rm b}$, in such starbursts is limited to
log~[T$_{\rm b}~$(K)]~$\lesssim$ 5 \citep{conet91}.
Thus it is necessary to show that T$_{\rm b}$ exceeds this limit before             
accretion onto a black hole can be claimed as the power source. 

From the observational perspective, \citet{hec80} showed that LINER nuclei 
tend to be associated with a compact radio source, and compact,
flat-spectrum radio cores are known to be present in many
`normal' E/S0 galaxies \citep{sadet89,wrohee91,sleet94}.
Flat-spectrum radio cores are, however, uncommon in normal spirals or 
Seyfert galaxies \citep{ulvwil89,vilet90,sadet95}, though \citet{ulvho01a} have
pointed out that low-luminosity Seyferts have a higher incidence of
flat-spectrum cores than more powerful Seyferts.

How does one distinguish accretion-powered LLAGNs from LLAGNs powered by hot 
stars or supernova shocks? Broad \ha\ lines and bright
unresolved optical or UV sources are ambiguous indicators because they can 
all be produced in starburst models \citep[see][]{teret92}.
Searches for broader polarized \ha\ emission are currently feasible in 
only a few of the brightest LLAGNs \citep{baret99}. X-ray emission from the nucleus
can be confused with sources in the bulge or disk of the galaxy, even at
\textit{Chandra}'s sub arcsecond resolution.
Further, all these indicators may be affected by viewing geometry,
obscuration, and the signal-to-noise of the observations, the last problem 
being exacerbated by the low optical, UV, and X-ray luminosities of LLAGNs
and the need to subtract the starlight. 
The radio regime, however, offers several advantages. Gigahertz (cm wave)
radiation does not suffer the obscuration that affects the UV to infra-red. 
Also, at tens of gigahertz the problems of free-free absorption can be 
avoided in most cases. Finally, high resolution, high sensitivity 
radio maps can be routinely made with an investment of less than an hour
per source at the Very Large Array{\footnote{The VLA and VLBA are operated by 
the National Radio Astronomy Observatory, a facility of the National Science 
Foundation operated under cooperative agreement by Associated Universities, Inc.}}
\citep[VLA;][]{thoet80} and
the Very Long Baseline Array \citep[VLBA;][]{napet94}. 
At their resolutions of $\sim\,$100~milli-arcsec (mas) and $\sim\,$1~mas, 
respectively, it is easy to pick out the AGN, since any other radio emission
from the galaxy is usually resolved out.

Closely related to these theoretical and observational issues are the
increasing number of accurate mass determinations for 
``massive dark objects'' (MDOs) in nearby galactic nuclei. 
Indeed, dynamic signatures 
of nuclear black holes are being found in almost every nearby galaxy 
studied. Mass estimates from high-resolution stellar-, maser-, or gas-dynamics 
are probably accurate to a factor of two \citep{ricet98,gebet00}.
Slightly less reliable (factor $\sim\,$2-5) estimates of M$_{\rm MDO}$ for large
samples are possible using the tight correlation between central stellar velocity
dispersion of the galaxy bulge ($\sigma_c$) and M$_{\rm MDO}$ \citep{merfer01,gebet00}.
In this paper we assume that the few~$\times$ 10$^{5-9}\,$\Msun\ 
MDOs traced by dynamic methods are black holes, and interchangeably use the 
terms MDO and black hole.
With the new MDO results the degeneracy between black hole mass and accretion 
rate when accounting for radiated luminosity is removed. 
Given AGN spectral energy densities and black hole
masses in large samples of nearby AGNs, one can attempt to
isolate factors which determine accretion rates and luminosities. 

A weak correlation between the nuclear centimeter radio luminosity 
and the mass of the nuclear black hole has been found by several authors, e.g. 
\citet{fraet98}, \citet{yibou99} and \citet{dimet01}.
This result has been used as evidence for the existence of an ADAF-like 
inflow since, in such a flow, the radio luminosity is predicted to scale with the
black hole mass and the accretion rate \citep[e.g.][]{mah97}.
Two significant weaknesses here are the low resolutions of the radio 
data, which allows significant contamination from other processes in the bulge,
and the small sample sizes.
Our 150~mas resolution radio observations of a large number of nearby LLAGNs 
considerably increase the number of LLAGNs with reliable black hole mass
estimates \textit{and} high resolution radio observations, allowing 
a better test of the relationship between the two quantities.

In the following sections we first define the samples used in the paper
(Sect.~\ref{secsample}), and report on new VLA and VLBA observations which 
complete our survey of the distance-limited (\d19) sample of LLAGNs 
(Sects.~\ref{secnewobs} and \ref{secnewres}). Observational
results on radio-variability in LLAGNs are also presented in Sect.~\ref{secnewres}.
In Sects.~\ref{secdetrate} to ~\ref{secgalbulge} we combine the new observations 
with those reported 
in the earlier papers in this series, and  present comprehensive results on the 
radio properties of LLAGNs in the distance-limited (\d19) sample, and their 
implications for the central engines in LLAGNs. 
In Sect.~\ref{secoverallmdo} we study the relationship between 
nuclear radio power and black hole mass, both for the distance-limited LLAGN sample
and for the larger sample of nearby bright galaxies (most of them
LLAGNs) defined in Sect.~\ref{secsample}. 
The results are followed by a brief discussion 
(Sect.~\ref{secdiscussion}) on the nature of the central engines in LLAGNs.

\section{Sample Selection}
\label{secsample}
\subsection{The LLAGN Sample}
\begin{figure*}
\resizebox{\textwidth}{!}{
\includegraphics{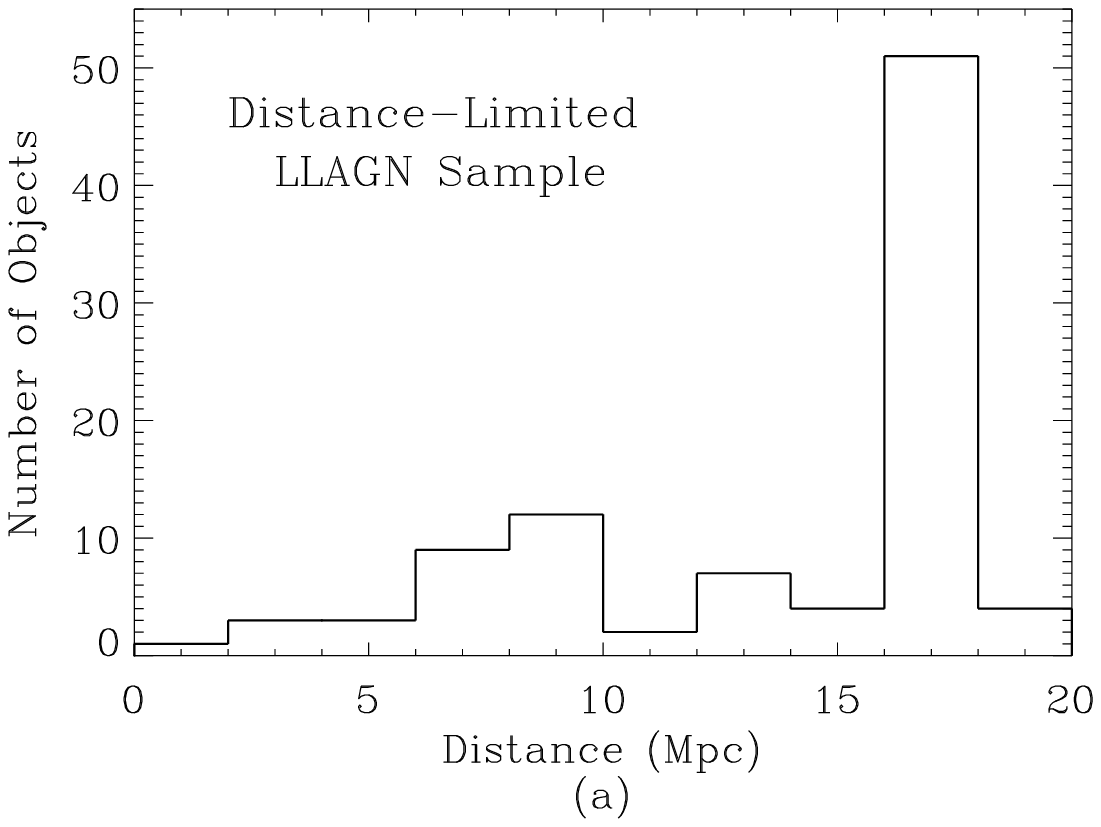}
\includegraphics{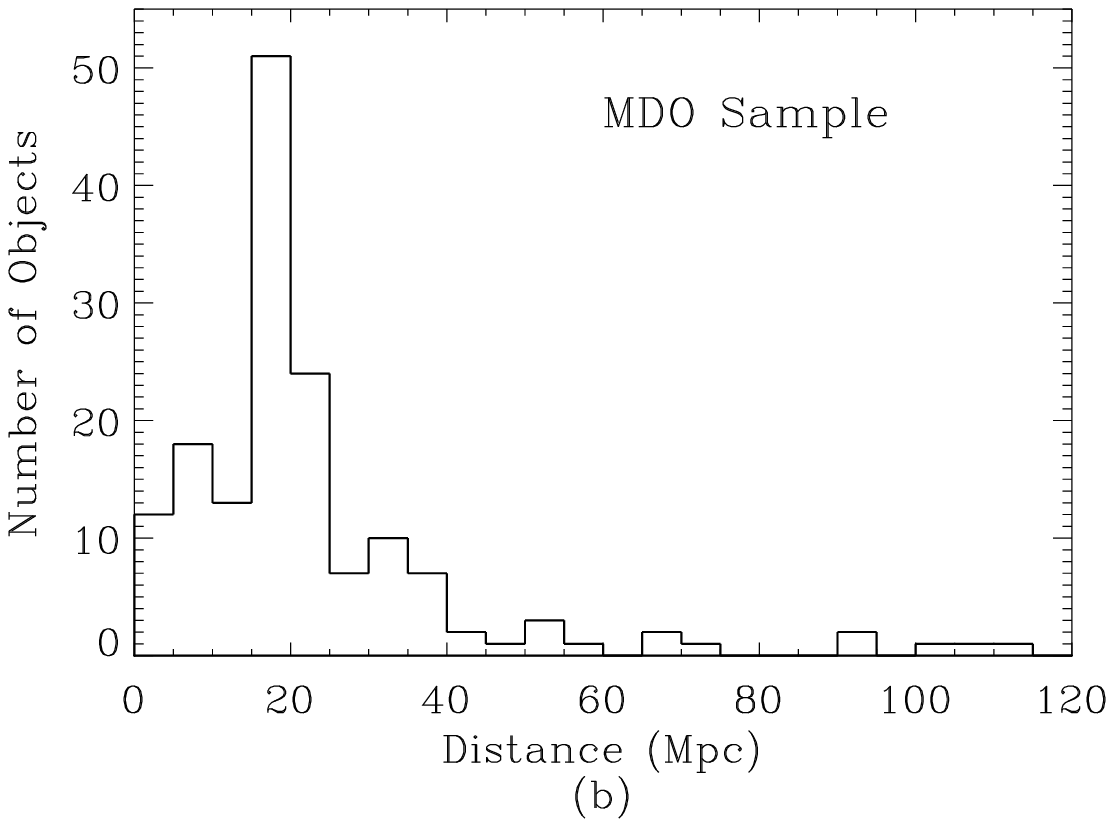}}
\caption{
Histograms of the galaxy distance for the 
\textbf{(a)} distance-limited (\d19) LLAGN sample, and
\textbf{(b)} MDO sample.
Virgo cluster members create the peak at D$\,\sim\,17\,$Mpc.}
\label{figdist}
\end{figure*}

The results in this paper, and the new observations reported here, are
based on LLAGNs selected from the Palomar spectroscopic survey of all
($\sim\,$470) northern galaxies with B$_{\rm T} <$~12.5~mag \citep{hoet95}. 
Almost half of all galaxies in the Palomar survey host some form of LLAGN 
\citep{hoet97b}. 
Spectroscopic parameters (including activity classification) of 403 galaxies in the 
Palomar sample have been presented in H97a, and from these
we have chosen to study the 96 nearest (\d19) LLAGNs 
-- the ``distance-limited LLAGN sample'' --
at 2~cm with the VLA at $\sim\,$0\farcs15 resolution. 
The distribution of galaxy distance for the distance-limited LLAGN sample is 
shown in Fig.~\ref{figdist}a.
Clearly, biases arising from the distance distribution are expected to be small within 
this sample.  Observations at 2~cm of 33 LLAGNs in the sample (taken in 
1996 October) were presented in \citet{naget00}.
New 2~cm observations of 72 LLAGNs in the sample are presented 
here (10 LLAGNs are in common between this work and Nagar et al. 2000,
and 1 LLAGN [IC~356] was not observed).

Our VLA observations of the 95 LLAGNs resulted in the detection of
31 nuclei. Sixteen of these have core flux 
S$^{\rm VLA}_{\rm 2cm} >$ 2.7~mJy, and all sixteen cores have a 
flat radio spectrum{\footnote{In view of the non-simultaneity of the 
radio data used to derive the spectral index, we use $\alpha~=\,-$0.3 
instead of $-$0.5 as the cutoff between a steep and flat spectrum.}}
($\alpha~\geq$ $-$0.3, S$_\nu~\propto \nu^\alpha$).
These sixteen objects - the ``distance- and radio-flux-limited LLAGN sample'', 
listed in Table~\ref{tabll16}, were chosen for further study at multiple 
frequencies with the VLA and at higher resolution with the VLBA.
Multifrequency VLA observations of the nuclei in Table~\ref{tabll16} 
(Nagar, Wilson, \& Falcke 2001) and VLBA observations of roughly half of the 
nuclei in Table~\ref{tabll16} have already been published (see column 5), 
while VLBA observations of the remaining seven nuclei in Table~\ref{tabll16} 
are presented in the present paper.

\subsection{The ``MDO'' Sample}
\label{secmdosample}
In order to investigate the relationship between MDO mass and nuclear radio 
luminosity, we have expanded our distance-limited (\d19) LLAGN sample as follows.
We include 73 additional Seyferts and LINERs (at D$\,>\,19\,$Mpc) from the list
of LLAGNs in H97a; these have also been observed at 2~cm and 150~mas resolution 
(Nagar et al. 2000, Nagar et al., in preparation). Also, we include an 
additional 31 galaxies which have reliable MDO mass estimates published in the 
literature. These include all galaxies in \citet{ricet98}, \citet{gebet00}, 
\citet{fermer00}, and \citet{merfer01}, and individual galaxies from 
\citet{shiet00} and \citet{bowet00}.  
Of the 200 galaxies, 154 have all three of the following: an estimated MDO mass 
(either directly from dynamic measurements or estimated from $\sigma_c$), a core 
radio flux, and a bulge luminosity. 
The distribution of galaxy distance for these 154 galaxies is 
shown in Fig.~\ref{figdist}b. 

\section{New Observations and Data Reduction}
\label{secnewobs}

\subsection{VLA Observations and Data Reduction}
\label{obsvla}

Seventy two LLAGNs at \d19 were observed at 2~cm (15~GHz) with the VLA.
The bulk of the new observations was carried out on 1998 March 3 and 1998 May 21 
while the VLA was in A-configuration \citep{thoet80} and on 1998 June 21 
while the array configuration was being changed from A- to B-configuration.
Data taken during the latter period have more or less the same resolution as 
A-configuration data as the outermost antennas of the array had not yet been moved.
Six LLAGNs (IC~239, IC~1727, NGC~428, NGC~660, NGC~1055, and NGC~1058)   
were observed on 1998 July 31 while the array was in B-configuration and
two LLAGNs (NGC~3941 and NGC~7177) were observed on 1998 September 5 and 
September 10, respectively, while the array was in A-configuration.
Most of the data (all data taken before July 1998) were calibrated and mapped by 
HF and preliminary results have been presented in \citet{falet99}. 
The remaining data were calibrated and mapped by NN.

Data were calibrated and mapped using AIPS, following the 
standard reduction procedures as outlined in the AIPS 
cookbook\footnote{available online at www.nrao.edu}. 
Observations of 3C~286 and 0404+768
(for which we used a 2~cm flux-density of 1.46~Jy) were used to set the 
flux-density scale at 2~cm. 
For sources stronger than about 3~mJy, we were able to iteratively self-calibrate 
the data so as to increase the signal-to-noise ratio in the final map. 
The typical root mean square (r.m.s.) noise in the final maps was 0.2~mJy and 
we take the detection limit of the survey to be 1~mJy (i.e. 5$\,\sigma$).
The resolution of the final maps was $\sim\,$0{\farcs}15 for the 
A-configuration maps and $\sim\,$0{\farcs}4 for the B-configuration maps. 

The 1$\sigma$ error in flux bootstrapping (i.e. setting the flux density 
scale relative to the flux calibrators) is expected to be roughly 2.5\%. 
Elevation dependent effects (atmospheric opacity and varying antenna 
performance with elevation) lead to additional errors when measuring the
fluxes of some sources (see Sec.~\ref{secvar}). 
Such elevation dependent effects are expected to be small for most sources 
as the flux calibrator sources were observed at elevations 
45{\degr} to 50{\degr} and most sources were observed at elevations between
40{\degr} and 60{\degr}. We therefore did not correct for elevation dependent
effects, but take them into account when estimating the error in the
flux measurement.

\subsection{VLBA Observations and Data Reduction}
\label{vlbaobs}
Nine of the sixteen nuclei listed in Table~\ref{tabll16} have been previously 
observed with the VLBA or VLBI (see column 5), and new VLBA observations of 
the remaining seven are presented here. We also observed two additional 
LLAGNs (with D$\,>\,$19~Mpc) from the Palomar sample. The first, NGC~2655, has
S$^{\rm VLA}_{\rm 2cm} >$ 2.7~mJy, but does not have a flat-spectrum
radio core \citep{naget00}. The other, NGC~3147 has both
S$^{\rm VLA}_{\rm 2cm} >$ 2.7~mJy and a flat-spectrum radio core
\citep[Nagar et al. 2000, see also][]{ulvho01b}.

These nine LLAGNs were observed at 6~cm (4.9 GHz)
in a 10 hour VLBA run on April 1, 1999.
Each source was observed at two different hour angles in order to obtain
good $(u,v)$-coverage. At each hour angle, a cycle of 6.5 min on source +
1 min  on a nearby phase calibrator was repeated four times. The total
integration time on each source was therefore 52 minutes. 
The ``fringe finder'' sources 4C~39.25 and 3C~345 were also briefly observed,
and a 7 minute observation of J1642+6856 was used to check (and correct) the 
amplitude calibration of the VLBA.
The weather was fair (and dry) at all VLBA sites, except for occasional
wind gusts in excess of 35 mph at Kitt Peak. There were a few
intermittent tape problems at several antennas; all data points with tape 
weights less than 0.7 (on a scale of 0 to 1) were flagged as bad.
Data were calibrated using AIPS, following the standard procedures outlined 
in chapter 9 of the AIPS cookbook. 
Bad $(u,v)$ data were deleted before the phase solutions of the 
phase-calibrator observations were transfered to the galaxy data.

The sources were initially imaged using AIPS task IMAGR. 
For sources stronger than about 4~mJy, we were able to iteratively
self-calibrate and image the data so as to increase the signal-to-noise 
in the final map. The self-calibration and imaging cycles were done using 
the DIFMAP package \citep{she97}, and the final self-calibrated $(u,v)$ files 
were then imaged using the AIPS task IMAGR.
The peak flux-density of the source typically increased by a factor of 1.3 
during the self-calibration process. Therefore, for sources
weaker than 4 mJy, on which accurate self-calibration was not possible,
we have multiplied the {\it{peak}} detected flux-density by 1.3 as a crude
attempt to correct for atmospheric decorrelation losses. The r.m.s.
noise in the final
uniformly weighted images is typically 0.15~mJy to 0.2~mJy, and the 
resolution between 2~mas and 5~mas.

\section{Observational Results}
\label{secnewres}

\subsection{Results of the New VLA Observations}
\label{res96}

The results of the new VLA observations are listed in Table~\ref{tabll96} with 
columns explained in the footnotes. All six LLAGNs observed in B-configuration 
(0{\farcs}4 resolution) were undetected. All remaining LLAGNs have been observed
in A-configuration or at a resolution equivalent to that of the A-configuration 
(Sect.~\ref{obsvla}). Thus, the fluxes for all radio detected sources are 
measured at $\sim\,$0{\farcs}15 resolution.
The radio positions for the detected nuclei are limited by the positional 
accuracy of the phase calibrators, which is typically 2--10~mas, 
and by the accuracy of the Gaussian fit to the source brightness distribution, 
which depends on the signal to noise ratio of the source detection. 
The overall accuracy should typically be better than $\sim\,$50~mas.
We have compared the radio positions derived here with optical positions 
from Cotton, Condon, \& Arbizzani (1999), which were measured from the digital
sky survey with typical 1$\sigma$ accuracy 1{\farcs}5--2{\farcs}5 in
each of right ascension and declination. The results (column 6 of 
Table~\ref{tabll96}) show a good ($\leq$~2$\sigma$) agreement in most cases.

Only two nuclei in Table~\ref{tabll96}, NGC~3628 and NGC~4486 (M~87), show 
reliable extended structure in our 2~cm maps. The extended emission in the 
starburst galaxy NGC~3628 is from a known star-forming region \citep{caret90}
while that in NGC~4486 is from the well-known radio jet \citep[e.g.][]{junbir95}. 
The absence of extended emission in the other nuclei
is not surprising as the high resolution may resolve it out; in addition, such 
emission is expected to be weak at the high frequency observed. For a few sources, 
a Gaussian fit (with peak flux-density, major and minor axes as free parameters)
to the nuclear radio source brightness distribution resulted in a Gaussian size 
slightly larger than the beam size so that the peak flux-density was slightly smaller 
than the total flux; however, the deconvolved source sizes were much smaller than 
half the beam size, so we consider these sources as unresolved.
Thus, most of the detected 2~cm nuclear radio sources are compact at the 
0{\farcs}15 resolution (typically 15--25~pc) of our survey. The implied brightness 
temperatures for the 2~cm compact nuclei are typically T$_b$~$\geq$ 10$^{2.5-4.0}$~K.

Non-simultaneous spectral indices for the 2~cm detected nuclei have been
estimated by a comparison with radio data at other wavelengths from the 
literature (as indicated in column 12 of Table~\ref{tabll96}).
While the actual value of the spectral index is uncertain
given the resolution mismatch and the non-simultaneity of the 
observations, we can determine whether the core is 
flat spectrum ($\alpha \geq~-$0.3; S$_\nu \propto \nu^{\alpha}$),
or steep spectrum ($\alpha <~-$0.3), as noted in column 11 of the table.
We are confident of the reality of the flat spectra obtained for the 
following two reasons.
Since compact flat-spectrum radio sources are often variable, the use of 
non-simultaneous data at two frequencies can cause some intrinsically 
flat-spectrum radio sources to be misclassified as steep-spectrum sources.
However, since extended steep-spectrum radio sources are not
variable, the use of non-simultaneous data at two frequencies
should rarely cause intrinsically steep-spectrum radio sources to be 
misclassified as flat-spectrum sources. Also, since the resolution is better at 
the higher frequency, resolution effects will tend to steepen the measured spectrum 
if extended emission is present. 
Six of the 2~cm detected sources are not marked as flat-spectrum sources in the table.
Of these, three show extended radio emission from jets or lobes - NGC~4388 \citep{falet98},
NGC~4636 \citep{stawar86}, and  NGC~4472 \citep{ekekot78}, and one shows extended
emission from star-formation - NGC~3628 \citep{caret90}.
This extended emission may dominate over any flat-spectrum nuclear component within 
the beam of our radio maps.
For the two remaining sources, NGC~4293 and NGC~4419, VLA archive data at 6~cm 
(0{\farcs}4 resolution; peak flux-densities 4.1~mJy/beam and 11.8~mJy/beam, respectively)
and 20~cm (1{\farcs}2 resolution; peak flux-densities 13.0~mJy/beam and 28.3~mJy/beam, 
respectively) show that both have steep-spectrum radio cores which are more or 
less unresolved at both frequencies.

\subsection{Variability of the VLA-Observed Radio Emission}
\label{secvar}
Except for a few well-studied cases, e.g.\ NGC~3031 [M~81] 
\citep{hoet99,bieet00} and NGC~5548 \citep{wro00}, little is known about 
radio variability in Seyferts and LLAGNs.
In the course of our observations, we have observed several LLAGNs at
multiple epochs using the same telescope (VLA), telescope configuration
(A configuration), wavelength (2~cm), and data reduction methods 
\citep[Nagar et al., in preparation, this work]{naget00,naget01}.
This offers us a unique opportunity to study radio variability in a sample 
of LLAGNs.  Specifically, the 2~cm emission from all sixteen LLAGNs listed in 
Table~\ref{tabll16} was observed at two epochs - March, May, or June 1998
(this work), and September~1999 \citep{naget01}.
The 2~cm emission from some of these sixteen LLAGNs was also observed 
in October~1996 \citep{naget00}.
The 2~cm emission from several other LLAGNs was also observed at 
these epochs.
In addition, the 3.6~cm emission from the sixteen LLAGNs in 
Table~\ref{tabll16} was observed in September 1999 \citep{naget01} and 
December 2000 (Nagar et al., in preparation).

The errors in the measured fluxes come from five significant factors (see
    the VLA Observational Status Summary for details{\footnote{www.nrao.edu}}:
(a)~the accuracy of flux bootstrapping from a VLA `flux' calibrator. The 3.6~cm fluxes
    at both epochs were bootstrapped from observations of 3C~286. 
    The 2~cm fluxes were bootstrapped from observations of either 3C~286 and 
    0404+768 (for which we used a 2~cm flux-density of 1.46~Jy at all epochs). 
    All VLA-recommended flux bootstrapping procedures were followed and the 
    (1$\sigma$) errors due to this factor are estimated to be 1\% at 
    3.6~cm and 2.5\% at 2~cm. 
(b)~variation of antenna gain with elevation. 
    The gain of the VLA antennas at 2~cm and 3.6~cm is maximum at elevation 
    E$_0\,=\,$45{\degr}--50{\degr} and decreases (i.e. decreasing the 
    estimate of the observed flux) as the elevation increases or decreases 
    from E$_0$. 
    At both epochs, all 3.6~cm observations were made at elevations 
    35{\degr}--70{\degr}
    with the flux calibrators observed at elevation $\sim$40{\degr}.
    The antenna gain variation over this elevation range is expected to be
    $\leq$0.3\% and we therefore ignored this effect at 3.6~cm.
    The antenna gain variation at 2~cm is larger; it is less than 1.5\% over 
    the elevation range 35{\degr} to 65{\degr}, but increases to 
    $\sim$4\% at elevation 80{\degr}. 
    The 2~cm flux calibrators were observed at elevations between 40{\degr} 
    and 55{\degr} in all cases. 
    The variation of gain with elevation was corrected for only in the 
    reduction of the September 1999 data. 
    For the other 2~cm data we took this effect into account when 
    estimating the error of the flux determination.
(c)~atmospheric opacity. The atmospheric opacity at the VLA is typically 
    0.02 at 2~cm and $\leq$0.01 at 3.6~cm. At these opacities,
    atmospheric absorption will result in a source having an observed flux 
    which is $<\,$0.5\% larger when observed at elevation 70{\degr} than
    at elevation 45{\degr}. We did not measure the atmospheric opacity 
    during any of the observing runs except that of December 2000. From
    the weather conditions during the observing runs it is reasonable to 
    assume the opacities were not significantly higher than the typical 
    values quoted above.
    We did not correct for atmospheric opacity effects in the 2~cm and 
    3.6~cm data. Therefore, for sources observed at elevations below 
    45{\degr} we use the typical opacities above to estimate the 
    additional error in the measured flux;
    It is notable that in the 
    elevation range 45{\degr}-70{\degr} this effect works opposite 
    to the effect discussed in (b), thus decreasing the overall error.
(d)~the r.m.s. noise in the final maps, which was typically 0.1~mJy and 
    0.2~mJy at 3.6~cm and 2~cm, respectively. 
(e)~uncertainties intrinsic to the self-calibration process itself. \newline
In summary, factor (a) usually dominates so that the 1$\sigma$ flux 
calibration errors are typically about 1\% and 2.5\% for 3.6~cm and 2~cm 
observations at the VLA, respectively. While we have taken some effort
to estimate the flux errors, other factors, e.g.  variability in the flux 
calibrators, may cause the errors to be higher than estimated here, 
especially at 2~cm.

\begin{figure*}
\resizebox{\textwidth}{!}{
\includegraphics{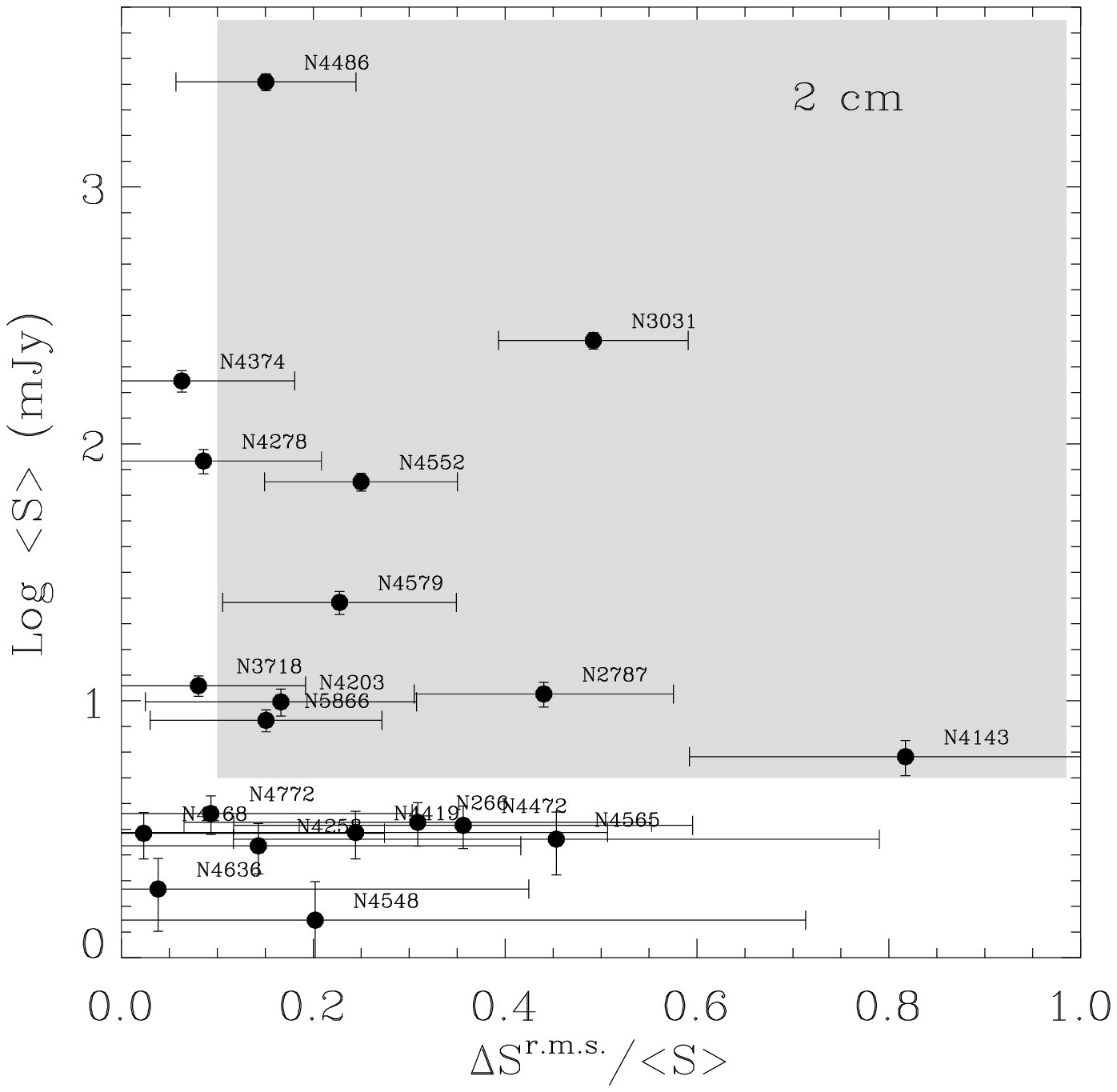}
\includegraphics{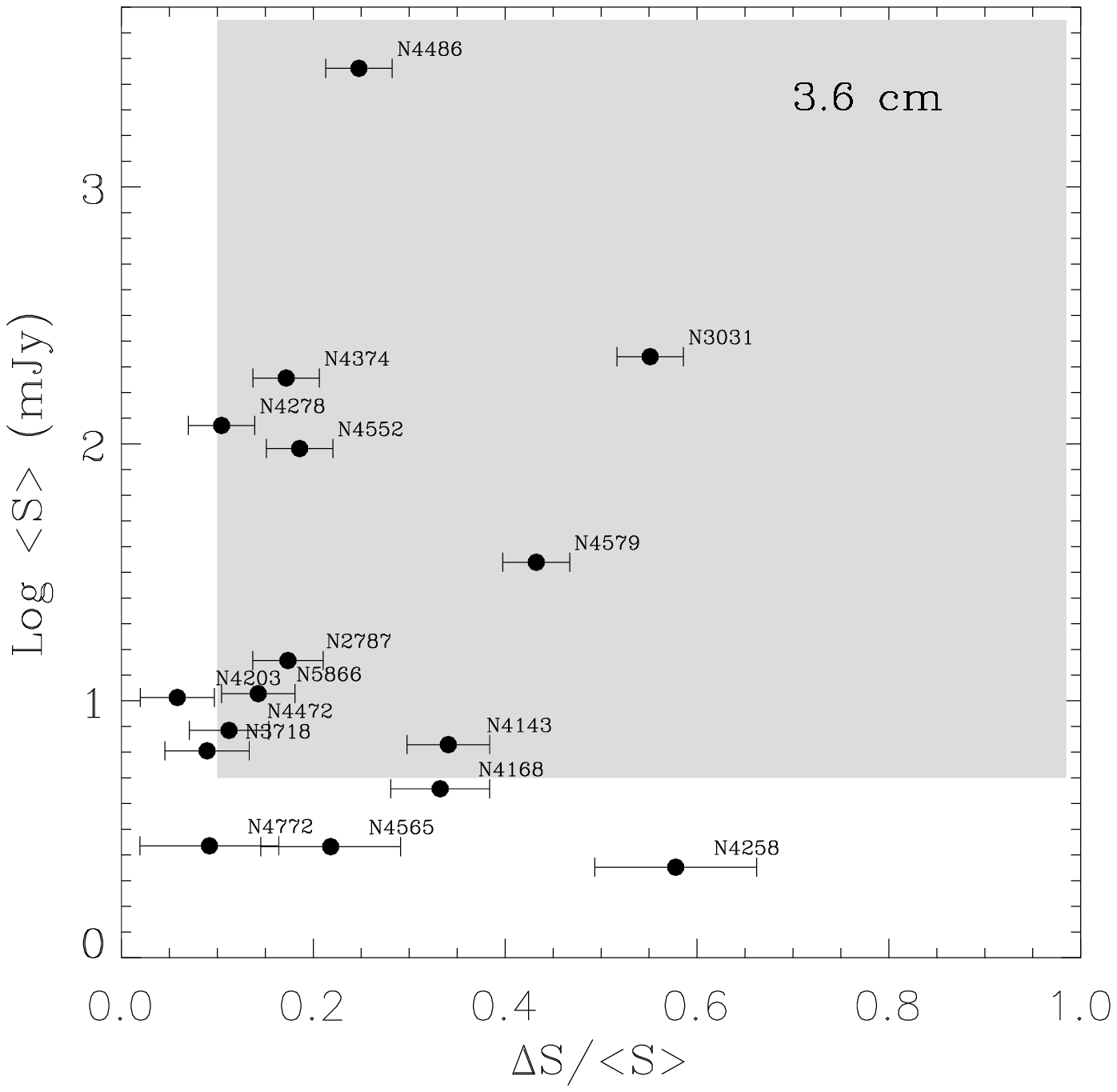}}
\caption{
 The inter-year variability of the nuclear \textbf{(left)}~2~cm flux,  and
 \textbf{(right)}~3.6~cm flux, in LLAGNs. 
 The $Y$ axis is the logarithm of the average measured flux, and the
 $X$ axis is the r.m.s. of the flux variation about the average flux 
 divided by the average flux.  For the 3.6~cm data, and a few 2~cm 
 datapoints, only two epochs are available and for these 
 cases $\Delta$S represents the difference of the two flux values.
 Estimated $\pm\,2\,\sigma$ errors in X and Y are shown.
 When error bars in Y are not shown, they are smaller than the plot symbol.
 Points outside the shaded area are likely to be dominated by
 errors in flux-calibration and/or radio imaging (see text); these points
 are therefore not reliable.} 
\label{figtimevar}
\end{figure*}

\begin{figure*}
\resizebox{\textwidth}{!}{
\includegraphics{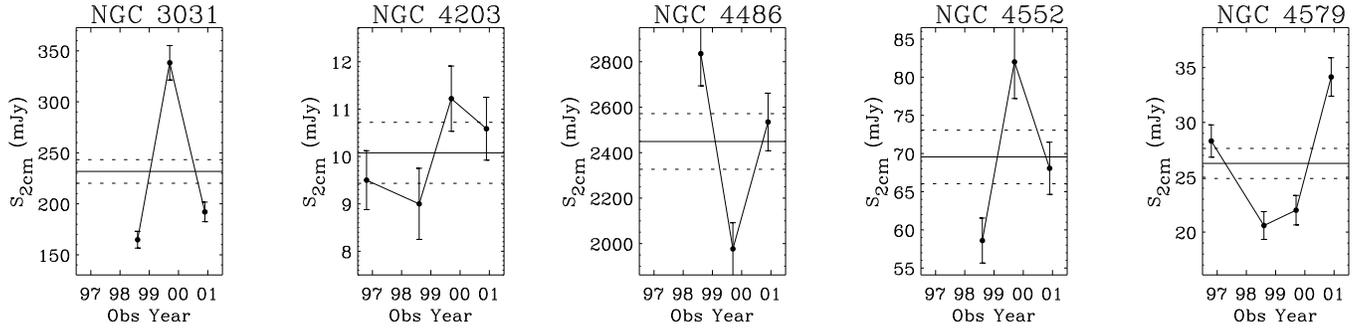}}
\caption{
 2~cm light curves for five of the LLAGNs in Fig.~\ref{figtimevar}.
 The 2~cm flux at the last epoch (December 2000) has been obtained
 by scaling the previous epoch's 2~cm flux by the ratio of the 3.6~cm fluxes,
 which were obtained on both epochs. Estimated $\pm\,2\,\sigma$ error bars 
 (from flux calibration and mapping errors) are shown for each datapoint.
 In each panel the horizontal solid line shows the average flux over the 
 observed epochs and the dashed lines show the estimated 
 $\pm\,2\,\sigma$ errors for the average flux value.}
\label{lightcurve}
\end{figure*}

The inter-year variability for the LLAGNs with multi-epoch
observations is shown in Fig.~\ref{figtimevar}.
We have plotted $<$S$>$ versus $\Delta$S/$<$S$>$, where $\Delta$S 
is the r.m.s. flux variation and $<$S$>$ the average flux (for nuclei with 
only two epochs of measurements, $\Delta$S represents the difference between
the fluxes).
The grey shading delineates the region within which observed variability can
be considered reliable. The lower cutoff to the shaded region is at 5~mJy. 
Sources with flux lower than this could not be accurately self-calibrated, 
which results in larger than usual errors in their flux determination.
The left cutoff results from uncertainty in the flux-calibration. This
uncertainty coupled with the r.m.s. noise in the radio maps could lead to
values of $\Delta$S/$<$S$>$ which are $\sim\,$10\%, even if the source
did not vary. As an alternate illustration of the flux variability we have
plotted the 2~cm `light curves' for five of the nuclei in Fig.~\ref{lightcurve}.
In Fig.~\ref{lightcurve} the data for the first three epochs are from our 2~cm 
observations. The data for the last epoch shown (December 2000) has been 
obtained by scaling the previous epoch's 2~cm flux by the ratio of the 3.6~cm 
fluxes which were determined at both epochs.

As can be seen in the Figures, we find significant variability in almost half
the sources.  While the 2~cm flux values potentially suffer from significant
flux calibration and elevation-dependent errors, the 3.6~cm flux-calibration
is quite reliable. The result is made more credible by earlier 
observations of variability in NGC~3031 \citep{hoet99,bieet00} and
NGC~4486 \citep{junbir95}. 
The large periods between the different epochs do not allow for detailed 
investigation of the variation, and at this point we are unable to find any 
clear correlation between the amount of variability and the other properties 
of the LLAGNs. Future higher time-resolution, multifrequency data will be
useful in this context.

\subsection{Results of the New VLBA Observations}
\label{resvlba}

All 7 objects from Table~\ref{tabll16} which were newly observed with 
the VLBA at 6~cm were clearly detected in initial maps
(i.e. without any form of self-calibration).
Of the two other (D~$>$ 19~Mpc) LLAGNs observed, NGC~3147 was detected, but 
NGC~2655 was not detected at a 10$\sigma$ limit of 0.76 mJy. The non-detection of
NGC~2655 is not surprising as this object has a steep-spectrum radio core at arcsec 
resolution \citep{naget00}.
Of the detected nuclei, mas-scale radio cores were known to exist in
NGC~4374 and NGC~4552 \citep[e.g.][]{jonet81}.
The results of the new VLBA observations are listed in Table~\ref{tabvlbanew}, 
with columns explained in the footnotes. 
The source positions are referenced to the positions of their respective 
phase-calibrators. All phase calibrators were drawn from the VLBA calibrator 
survey and the positions we used at the time of observation were 
expected to be accurate to $\pm$10 mas in each of R.A. and Dec. 
The phase calibrator positions have now been determined to an accuracy
of 1~mas or better \citep{beaet01} and the new positions are available on the
web{\footnote{www.nrao.edu}}. We therefore updated the positions of the sources
to reflect the new accurate phase calibrator positions.
The other factor contributing to the
position uncertainty is the accuracy of the Gaussian fit to the 
source, which should typically be better than 2~mas. 
Thus the overall accuracy of the positions listed in Table~\ref{tabvlbanew} 
should be about 2~mas or better. The implied brightness 
temperatures were calculated using the formula given in \citet{falet00};
the results are in the range 10$^{6.8}$ to 10$^{9.5}$ K. Since most
of the sources are unresolved, these values are lower limits to the true 
brightness temperatures.

\begin{figure*}
\resizebox{\textwidth}{!}{
  \includegraphics{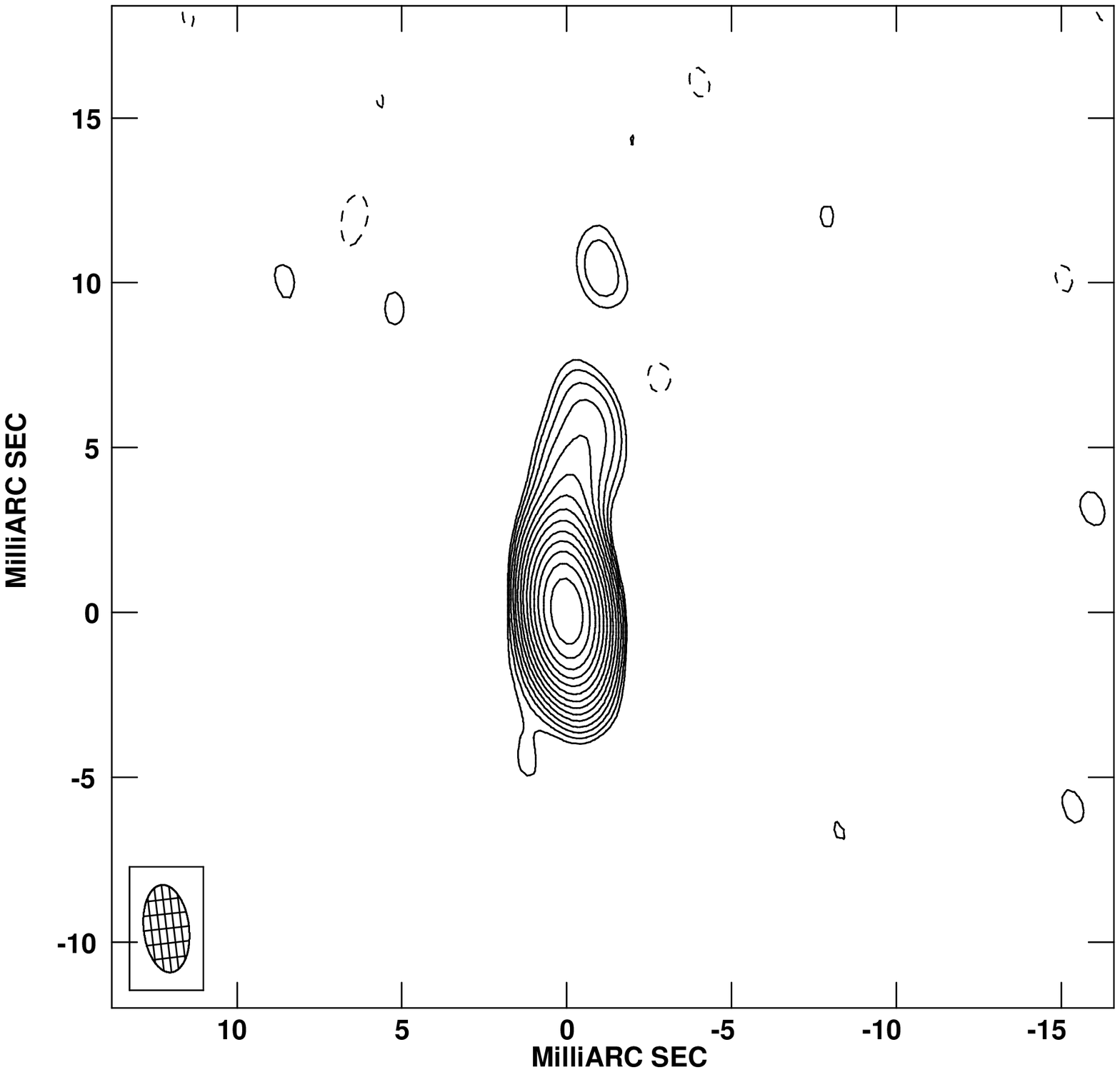}
  \includegraphics{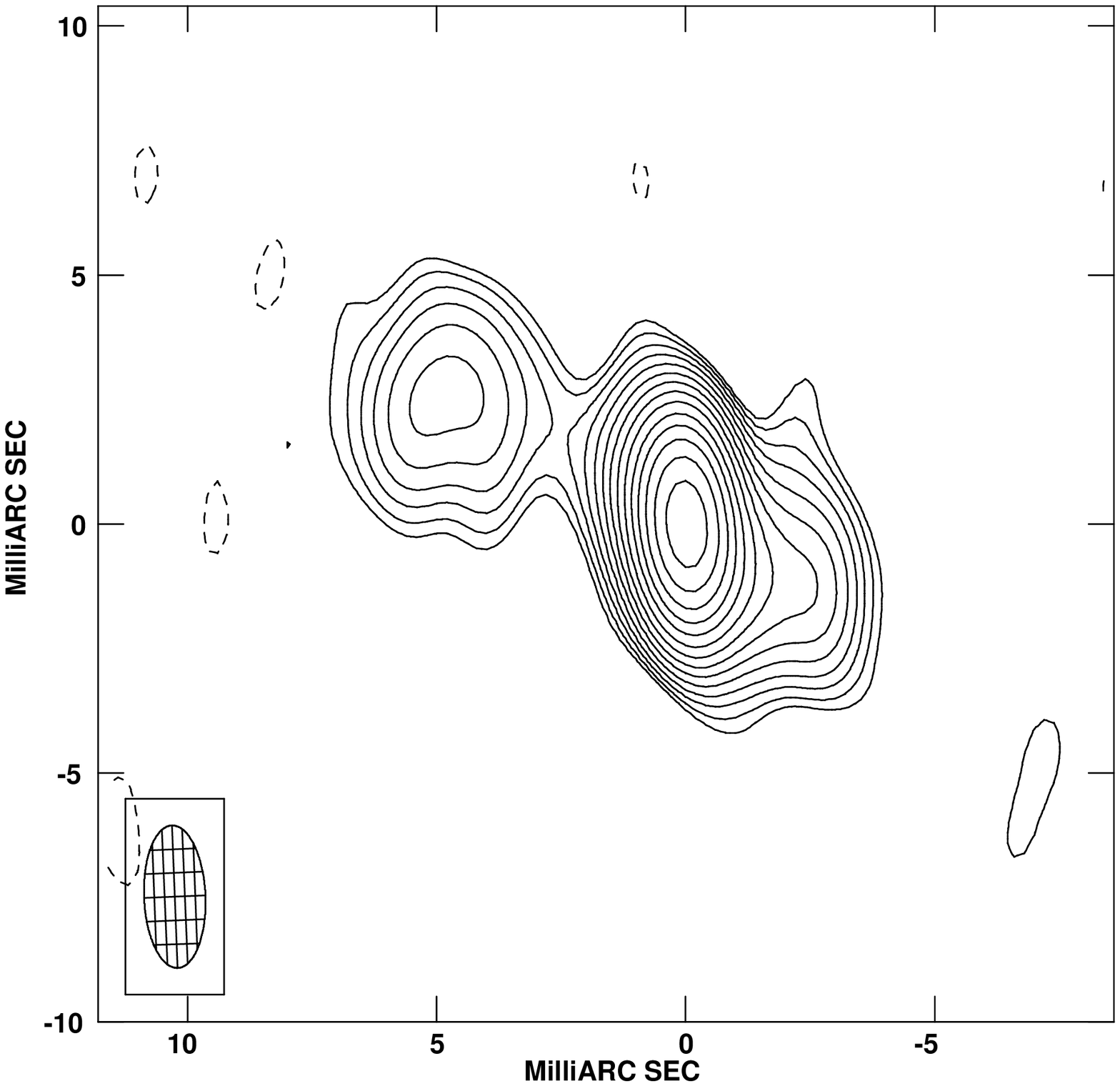}}
\caption{
 5~GHz (6~cm) VLBA maps of NGC~4374 \textbf{(left)} and 
 NGC~4552 \textbf{(right)}. 
 The contours are integer powers of $\sqrt{2}$, multiplied by the 
 $\sim\,3\,\sigma$ noise level of 1.2~mJy for NGC~4374, and by the
 $\sim\,2\,\sigma$ noise level of 0.8~mJy for NGC~4552.
 The peak flux-densities are 152.6~mJy/beam and 93.8~mJy/beam, 
 respectively.}
\label{vlbamaps}
\end{figure*}

Among the newly observed sources, 
the two with the highest peak flux-density at 6~cm, NGC~4374 (M~84) and 
NGC~4552 (M~89), clearly show extended structures (Fig.~\ref{vlbamaps}) 
suggestive of  parsec-scale ``jets.'' The pc-scale extension in NGC~4374 
\citep[detected earlier by][]{wroet96} is aligned 
with the larger kpc-scale FR-I radio lobes \citep[e.g.][]{laibri87}, while
the twin pc-scale extensions in NGC~4552 are, to our knowledge, the first 
time such structure has been detected in this galaxy.
As pointed out in \citet{naget01}, of the sixteen objects in Table~\ref{tabll16}
the five with the highest 6~cm peak 
flux-densities at mas resolution --
NGC~3031 \citep[M~81;][]{bieet00},
NGC~4278 \citep{jonet84,falet00},
NGC~4486 \citep[M~87;][]{junbir95},
NGC~4374 \citep[M~84;][this work]{wroet96},
and NGC~4552 (M89; this work) --
all have one- or two-sided pc-scale extensions, suggestive of jets. 
For these sources the peak flux-density of 
the extended emission is typically 10\%--25\% of the peak flux-density of 
the core. The signal-to-noise ratios of the core detections for the other 
eleven sources listed in Table~\ref{tabll16} range between 10 and 50. The 
relative weakness of most of these sources does not allow very good 
self-calibration.  This weakness, coupled with the low signal-to-noise expected 
for any extended component (10\%--25\% of a 10$\,\sigma$--50$\,\sigma$ 
detection) may provide an explanation for the non-detection of weak
extended structure in most of these sources. In fact, very deep VLBA 
observations of NGC~4258 reveal a sub-pc jet \citep{heret97}. 

\section{Detection Rates of Radio Cores in the Distance-Limited Sample}
\label{secdetrate}
\subsection{Parsec-Scale Radio Cores (VLBA Detections)}
\label{secdefinite}
The new VLBA results confirm that \textit{all} LLAGNs at \d19
with a 2~cm compact flat-spectrum core (as determined from
VLA observations), i.e. 17\% $\pm$4\% of all LLAGNs at \d19, 
have mas-scale radio cores with brightness-temperature $\gtrsim\,10^8\,$K.  
It is notable that this sample of LLAGNs with ultracompact radio cores is
almost equally divided between Seyferts (6 of 16) and LINERs (8 of 16, see
Table~\ref{tabll16}).
Nuclear starbursts can have a maximum brightness-temperature of 
$\sim\,10^{4-5}\,$K \citep{conet91} while the most luminous known radio 
supernova remnants \citep[e.g.][]{colet01} would have brightness 
temperatures $\leq\,10^{7}$~K even if they were $\leq$~1~pc in extent.
As argued in \citet{falet00}, if the core radio emission is attributed to
thermal processes, the predicted soft X-ray luminosities of LLAGNs would be 
at least two orders of magnitude higher than observed by \asca \citep{teret00}
and \chandra \citep{hoet01}.
Finally, as discussed in \citet{ulvho01a}, single SNRs \citep{colet01} 
or a collection of SNRs \citep{nefulv00} would have radio spectral indices 
$\alpha\,\sim\,-$0.7 to $-$0.4 rather than the values $\alpha\,\sim$ $-$0.2 
to 0.2 seen in the VLBA-detected LLAGNs \citep{naget01}.
Thus, the only currently accepted paradigm which may account for the mas 
radio cores is accretion onto a supermassive black hole. In this case,
the mas-scale radio emission is likely to be either emission 
from the accretion inflow \citep{naret98} or synchrotron emission from 
the base of the radio jet launched by the accreting supermassive black hole 
\citep{falbie99,zen97}.

Fifteen of the sixteen definite AGNs (Table~\ref{tabll16}) are either
(a)~type~2 LLAGNs in massive (M$_{\rm B}$(total) $\leq\,-20$) ellipticals
    (four galaxies),
    or
(b)~type~1 LLAGNs (two ellipticals and nine non-ellipticals).
The remaining galaxy, NGC~5866, is a transition nucleus in an S0
galaxy with an ambiguous detection of broad \ha\ (H97c); the ambiguity 
is not unexpected as the dilution by \hii\ regions in transition nuclei makes
it more difficult to determine the presence of broad \ha\ than in 
``pure'' LINERs and Seyferts (see H97c). 
NGC~5866 may, therefore, belong with the other nine type~1 non-ellipticals.

\subsection{100-Parsec-Scale Radio Cores (VLA Detections)}       

We have completed observations of 95 of the 96 LLAGNs in the distance-limited 
sample with the VLA; only IC~356, a transition nucleus, remains unobserved by
us. IC~356 can be considered as a non-detection for our purposes since a 
significantly lower resolution (1{\arcmin}) flux measurement gives an 
upper limit of 2~mJy at 2.8~cm \citep{niket95}.

\begin{figure*}
\resizebox{\textwidth}{!}{
\includegraphics[bb=72 640 540 864,width=\textwidth,clip]{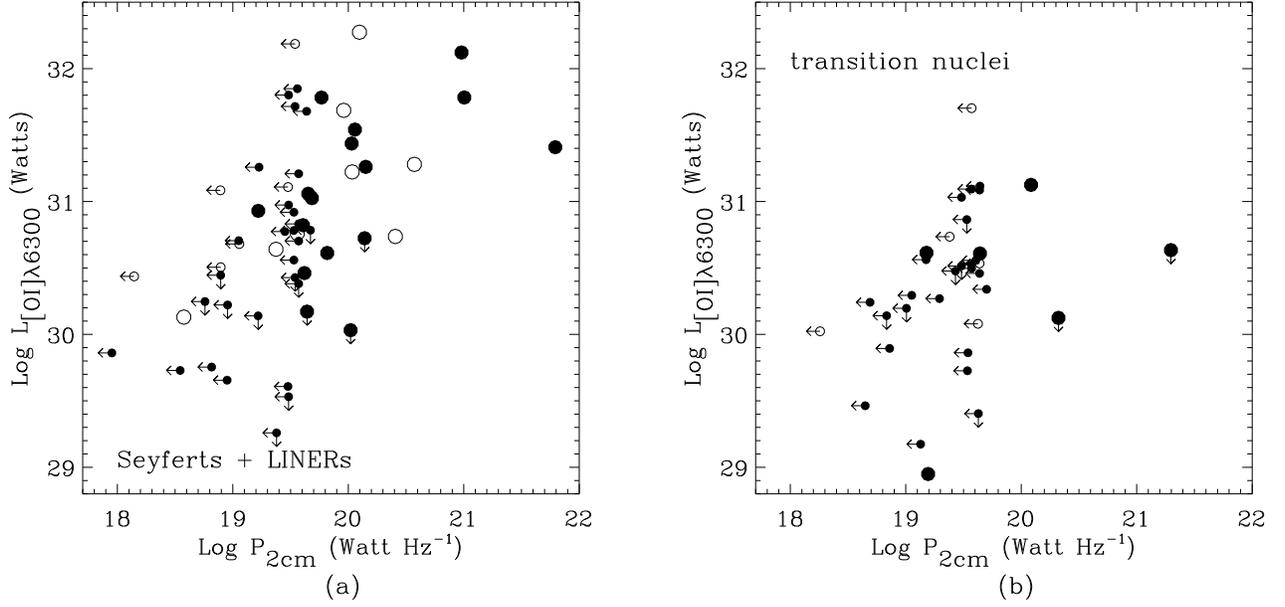}}
\caption{
 Relationship between [O~I]~$\lambda$6300{\AA} luminosity and 2~cm 
 core (150~mas resolution) radio power for 
 \textbf{(a)}~all LINERs and low-luminosity Seyferts, and 
 \textbf{(b)}~all transition nuclei, 
 in the distance-limited (\d19) sample.
 Large symbols are used for radio detections.
 Objects with non-photometric [O~I]~$\lambda$6300{\AA} flux measurements
 are shown as open circles.}
\label{radvso1}
\end{figure*}

When all 96 LLAGNs at \d19 are considered together, our VLA observations have
detected 9 of 23 (39\%) low-luminosity Seyferts{\footnote{NGC~185 is 
considered a non-detection since the radio source detected in this 
galaxy is probably extra-nuclear \citep{naget00}.}},
16 of 37 (43\%) LINERs, and 6 of 35 (17\%) transition nuclei, at a resolution
of $\sim\,$0{\farcs}15 and above a flux limit of $\sim\,$~1~mJy 
(Nagar et al. 2000; this work). Alternately, we can state that radio cores 
with luminosity P$^{\rm core}_{\rm 2cm}~\geq$ 10$^{20}$ W~Hz$^{-1}$
are found in 5 of 23 low-luminosity Seyferts, 8 of 37 LINERs, and 3 of 35
transition nuclei in the distance-limited sample.
The radio luminosities of the detected 2~cm cores lie between
10$^{18}$ and 10$^{22}$ W Hz$^{-1}$ (e.g. Fig.~\ref{radvso1}), similar to the
luminosities seen in `normal' E/S0 galaxies
(Sadler et al.\ 1989). It is notable, however, that a significant fraction of
the detected 2~cm compact cores are in spiral galaxies (Fig.~\ref{detvsnon}a).

\section{Correlations between Radio Power and Optical Emission-Line
         Properties in the Distance-Limited Sample}
\label{secrado1}

Emission-line fluxes and linewidths for the galaxies in the Palomar sample 
were typically measured over a 4{\arcsec}$\,\times\,$2{\arcsec} aperture 
centered on the galaxy nucleus (H97a). In this section 
we study the relationships between these nuclear emission-line properties and 
the core radio power measured at 150~mas resolution for LLAGNs in the
distance-limited (\d19) sample.

\subsection{Emission-Line Luminosity}

Correlations between the radio power and emission-line luminosity of active
galaxies are well known, and are found in both Seyfert \citep{debwil78}
and radio galaxies \citep{bauhec89}.
The \fulloiii\ line is usually used in assessing these correlations,
but this line is weak in LINERs and we use instead \fulloi, which
also has the advantage of being uncontaminated by emission from \hii\ regions. 
In our \d19 LLAGN sample the 2~cm radio power appears approximately proportional 
to the nuclear \fulloi\ luminosity (Fig.~\ref{radvso1}), though there is a range 
of about two orders in magnitude of emission-line luminosity for a given radio 
power. Testing the statistical significance of the correlation is
not straightforward as there are some uncertain values
(non-photometric data) of the \oi\ luminosities.
For simplicity, we treat the non-photometric flux densities as
photometric measurements when using the bivariate tests from the
ASURV statistical software package \citep[][hereafter Asurv]{lavet92}.
This approximation should not bias the results significantly as the 
uncertainties of the non-photometric emission-line fluxes are expected to be a 
factor of only $\pm$2 (H97a), which is small compared to the 3 orders of magnitude 
range of the \oi\ luminosity. When this is done, the Asurv tests indicate that 
the \oi\ luminosity is correlated with the 2~cm radio power at the 98.5\%--99.9\% 
significance level for the thirty-one 2~cm detected LLAGNs at \d19, at the 
99.9\% significance level for all Seyferts and LINERs at \d19, and at the 
99.99\% significance level for all 95 LLAGNs at \d19 which have been observed at 2~cm 
(see Table~\ref{tabstatll96}).
The correlation among the radio-detected nuclei is not just a result of the
elliptical galaxy nuclei having higher radio powers and emission-line luminosities 
than the non-ellipticals; among the 24 non-elliptical radio detections, the radio power 
and \oi\ luminosity are correlated at the 99.5\% significance (Table~\ref{tabstatll96}).

Interestingly, if we use the \fullhanii\ luminosity in place of
the \oi\ luminosity the data appear more scattered, and the 
significance of the correlation drops to 97\%-98\% for all radio 
detected LLAGNs and 89\%-93\% for the radio-detected non-elliptical LLAGNs
(Table~\ref{tabstatll96}). 

\subsection{Emission-Line Width}  
\label{secwidth}
The only emission line for which the full width at half maximum (FWHM) is reported 
in the Palomar survey results is \fullnii\ (H97a). 
We therefore use this line for assessing the correlations between linewidths in 
LLAGNs and other parameters.
For all radio detected nuclei considered together, all three of
the \fullnii\ FWHM linewidth, \fullnii\ luminosity, and 2~cm radio power
are correlated, with the significance of the correlations as follows:
linewidth and radio power 99.9\%, 
linewidth and \nii\ luminosity 99-99.5\%, and
\nii\ luminosity and radio power (97-99\%; Table~\ref{tabstatll96}). 
The sample of low-luminosity Seyferts studied by \citet{ulvho01a} also follows 
the above three correlations at a high significance level.
The \nii\ linewidths of LINERs and transition nuclei are generally 
larger for the earlier type galaxies (the Seyferts do not show such a clear
variation with morphological type). 
Also, the same early type galaxies tend to have higher radio powers though
not necessarily higher \nii\ luminosities.
Interestingly, the \nii\ linewidth does not show any clear dependence on the 
bulge or total magnitude of the galaxy among the LLAGNs with \d19.

\begin{figure*}
\resizebox{\textwidth}{!}{
\includegraphics{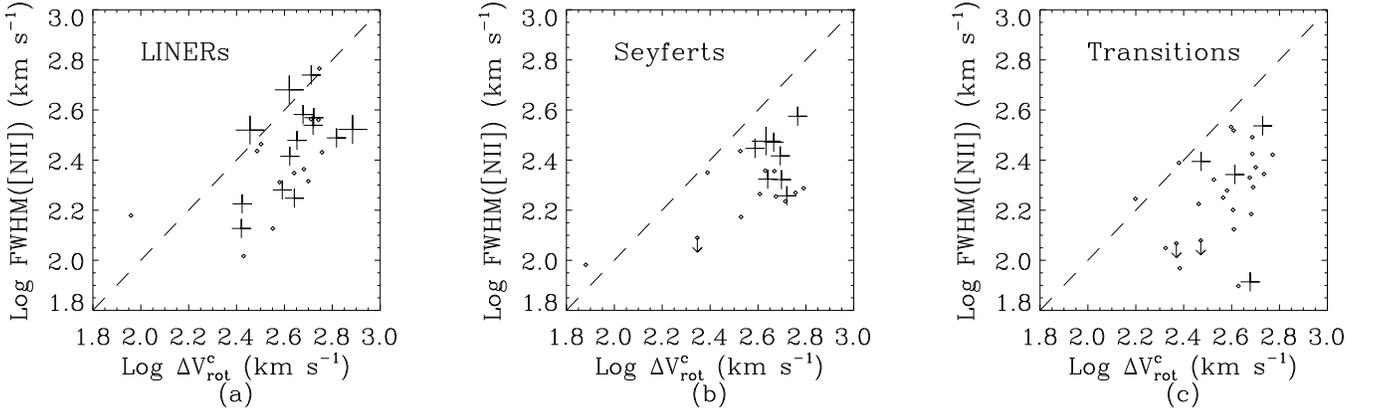}}
\caption{
 Relationship between the FWHM of the nuclear \fullnii\ emission-line 
 (as listed in H97a) and the inclination-corrected rotational velocity of
 the galaxy ($\Delta$V$^c_{\rm rot}$; as listed in H97a) for
 \textbf{(a)}~LINERs, \textbf{(b)}~Seyferts, and \textbf{(c)}~transition nuclei, 
 in the distance-limited (\d19) sample. 
 Radio-detected ellipticals are shown with large crosses, radio-detected 
 non-ellipticals with small crosses, and all radio non-detections with dots.}
\label{vrotvsfwhm}
\end{figure*}

\begin{figure*}
\resizebox{\textwidth}{!}{
\includegraphics{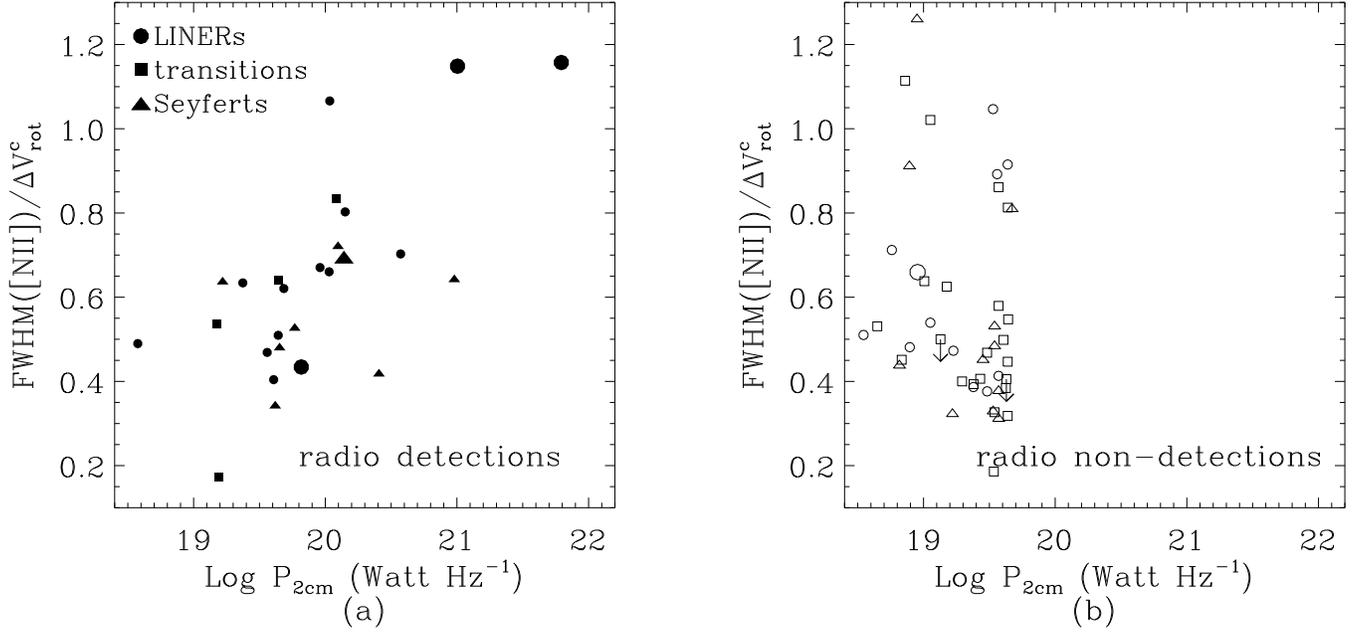}}
\caption{
 Relationship between the ratio of the FWHM of the nuclear \fullnii\ 
 emission-line to the inclination corrected rotational velocity of the galaxy
 and the 2~cm core (150~mas resolution) radio power for 
 \textbf{(a)}~radio detected nuclei and 
 \textbf{(b)}~radio non-detected
 nuclei in the distance-limited (\d19) sample. Large symbols are used for 
 elliptical galaxies and filled symbols for radio detections. 
 The symbols in the right panel thus represent upper limits to the radio power.}
\label{radvsratio}
\end{figure*}

That the galaxy rotation contributes to the \nii\ linewidth is clear:
the \nii\ linewidth is correlated with the corrected (for galaxy inclination)
rotational velocity of the host galaxy ($\Delta$V$^c_{\rm rot}$) for all
LINERs (99.7\% significance) and transition nuclei (97\%-98\% significance)
with \d19 (Fig.~\ref{vrotvsfwhm} and Table~\ref{tabstatll96}).
The significance of the above correlation for only the radio detected LINERs 
is smaller (82\%-95\%) perhaps due to the small number of objects.
More interestingly, there is evidence that the nuclear radio power 
is related to the ratio FWHM(\nii)/$\Delta$V$^{c}_{\rm rot}$, which
increases with increasing radio power (Fig.~\ref{radvsratio}). 
This correlation has a significance of 99.9\% if all radio detections are 
considered, and 98\%-99\% if only the non-elliptical radio detections are 
considered (Table~\ref{tabstatll96}). That this correlation may involve a 
direct physical significance is supported by the lack of a clear correlation 
between \nii\ width and bulge luminosity among the LLAGNs with \d19.
One immediate explanation for the correlation is that the radio source is 
responsible for ``stirring'' the \nii\ emitting gas as is probably the
case with powerful Seyferts \citep{wilwil80,whi92b}.
The converse, however, is not true. That is, a high ratio of
FWHM(\nii)/$\Delta$V$^c_{\rm rot}$ does not necessarily
imply a strong radio source: notice how the radio non-detections 
(Fig.~\ref{radvsratio}b) show a large variation in the y-axis. This is
one of the few suggestions that the radio non-detected nuclei are 
not simply lower radio-power versions of the radio detected nuclei.
In  both Figs.~\ref{vrotvsfwhm} and \ref{radvsratio}, the radio-detected 
nuclei NGC~4143, NGC~4168, NGC~4486, NGC~4552, and NGC~5866 are not plotted 
as these nuclei do not have values for the galaxy rotational velocity 
listed in H97a.
The only three radio detected nuclei with \nii\ linewidths in excess of 
$\Delta$V$^c_{\rm rot}$ (Figs.~\ref{vrotvsfwhm} and \ref{radvsratio}) are 
NGC~4278, NGC~4374 (both ellipticals with pc- to kpc-scale radio jets), and 
NGC~4772 which has a pc-scale radio core (Sect.~\ref{resvlba}).

\section{Comparison of the Radio Detected and Non-Detected LLAGNs in the
         Distance-Limited Sample}       
The sixteen LLAGNs observed at mas resolution and discussed in 
Sect.~\ref{secdefinite}
represent a flux-limited sample drawn from the 2~cm VLA-detected LLAGNs.
It is likely that some fraction of the VLA-detected radio cores which
have not yet been observed with the VLBA, and also some fraction of
the VLA-non-detected LLAGNs, also have high-brightness temperature cores 
(and thus accreting black holes).
These issues, crucial for determining the true incidence and properties of 
accreting black holes in nearby galaxies, are explored in the following 
sections. The fluxes and detection rates quoted in the following sections
are those from the VLA survey.

\subsection{LINERs and Low-luminosity Seyferts}
\label{seclinllsey}

The type~2 nuclei (broad \ha\ not detected) have a rather low VLA radio core
detection rate: 
only 1 of 11 Seyfert 2 nuclei and only 8 of 27 LINER 2 nuclei are 
detected.  The type~1 nuclei (broad \ha\ present) have a significantly higher 
detection rate: 
8 of 10 LINER 1.9s, and 8 of 12 Seyfert 1s (1.0s to 1.9s) are detected at 2~cm.
The difference between the detection rates of type~1s and type~2s increases 
when one considers the type classification more carefully. Of the 22 type~1 nuclei
in the distance-limited sample, a definite detection \citep[see][]{hoet97c} of broad 
\ha\ has been 
made in 17; the remaining five have probable detections of broad \ha.
Also, 6 of the 38 LINER type~2 and Seyfert type~2 nuclei in the distance-limited 
sample have ambiguous \citep{hoet97c} detections of broad \ha.
Of the definite type~1s, 13 of 17 (76\%) are detected in the radio. The only four 
non-detected definite type~1s are NGC~2681, NGC~4051, NGC~4395, and NGC~4639. 
Excluding type~2s with an ambiguous detection of broad \ha,
7 of 32 (22\%) type~2 LINERs and Seyferts are detected in the radio.

It is known that the radio powers of AGN are correlated with various parameters
of the host galaxy, such as the bulge luminosity \citep[e.g.][]{nelwhi96}
and possibly the nuclear luminosity at optical/UV wavelengths 
\citep[e.g.][]{debwil78,bauhec89}. It is, therefore, important to see if the 
preferentially higher detection rate of the type~1 nuclei over the type~2s is 
influenced by these effects. 
The single type~2 Seyfert detected in the radio, NGC~4472, is a massive
(M$_{\rm B}$(total) $\leq -20$) elliptical.
Of the eight type~2 LINERs detected in the radio, two (NGC~4374 and NGC~4486)
are massive ellipticals, one (NGC~2841) has an ambiguous identification
of broad \ha (H97c), four have very weak ($\sim\,3$-$7\,\sigma$)
radio detections and the last (NGC~4736, 1.9~mJy or $\sim10\,\sigma$,
distance 4~Mpc) is unusually nearby and would not be detected at the median
distance of the sample. Turning to the type~1s, 
the two non-detected LINER 1.9s are NGC~2681 and NGC~4438. 
The former has among the lowest bulge luminosity, emission-line 
luminosities and \nii\ FWHM in the LINER 1.9 sample, and the
broad \ha\ detection is doubtful in the latter (H97c). 
The four non-detected Seyfert type 1s are NGC~3982, NGC~4051, NGC~4395,
and NGC~4639. 
The broad \ha\ detection in NGC~3982 is doubtful (H97c) and 
the narrow \ha\ luminosity of NGC~4051 places it exactly on the 
operational cutoff between low-luminosity Seyferts and `classical' 
Seyferts, so it may have a different type of central engine from LLAGNs. 
A sub-mJy mas-scale nuclear radio source with brightness temperature
$> 2\times 10^6$, potentially an accreting massive black hole, has been 
found in NGC~4395 \citep{wroet01}. 
Such a weak radio core is expected in view of the extremely low bulge 
luminosity (M$_{\rm B}$(total)\,=\,$-$10.4) of this galaxy.
In summary, the radio-detected type~2s tend to be strong radio sources 
in high bulge luminosity galaxies or very weak radio sources, and of
the six type~1 nuclei not detected in the radio, two have 
a doubtful detection of broad \ha\ and two have unusually low bulge 
luminosities which may be the reason for their radio non-detection.
The preference for radio detected LLAGNs to be in type~1 nuclei
is the only highly significant difference
between the 2~cm detected and non-detected LINERs and Seyferts. 
Thus, it is most likely that the difference in radio power is indeed 
related to the AGN type.

In this context it is relevant to look at the detection rate of
the elliptical LLAGNs at \d19.
All four massive (M$_{\rm B}$(total) $\leq -20$) type~2 ellipticals 
(one of which has a transition nucleus) and all three type~1 ellipticals 
(with M$_{\rm B}$(total) values of
$-$19.1, $-$18.9, and $-$20.7) have detected radio cores.
Only three ellipticals (all type~2) remain undetected in the radio: NGC~185,
NGC~3379, and NGC~4494. 
NGC~185 is a nearby very low bulge mass 
(M$_{\rm B}$(total)\,= $-$14.9) elliptical and is 
thus expected to have a very low luminosity radio core; NGC~3379 
(M$_{\rm B}$(total)\,= $-$19.4) may actually be an S0 galaxy \citep{sta01};
NGC~4494 (M$_{\rm B}$(total)\,= $-$19.4) has a very low emission-line 
luminosity (log~[\Lha/W] $<$ 30.54; H97a). As noted above, the radio 
detection of the massive ellipticals would be expected if the
radio power scales with the bulge luminosity in LLAGNs 
(Sect.~\ref{secbulrad}). Among less massive
ellipticals it may be that the type~1 nuclei are preferentially
detected similar to the case for non-ellipticals.

Thus the demography of LLAGNs with VLA-detected radio cores is similar to 
that of the subset with VLBA-detected radio cores, i.e.  radio cores are 
preferentially seen in either massive type~2 ellipticals or in type~1 
LLAGNs. It is likely that we have detected 2~cm radio cores and broad \ha\ 
lines in only the more luminous LLAGNs, since the \fulloi\ luminosities of 
the Seyfert and LINER type~1 nuclei are higher than those of the type~2 
nuclei at the 99.9\% significance level. 
\citet{ulvho01a} arrived at a similar conclusion in their analysis of the
radio powers of $\sim\,45$ Seyferts drawn from the Palomar sample.
Further support for this idea comes from an apparent proportionality between 
core radio power and luminosity of the broad \ha\ component 
(Fig.~\ref{broadha}a) for all (20) type 1.9s in the Palomar sample for which
broad \ha\ luminosities were measured under photometric conditions 
\citep{hoet97c}. Twelve of the plotted nuclei are in the distance-limited 
sample; radio data for the remaining nuclei were taken from Nagar et al. 
(in preparation). The proportionality is unlikely to be due to distance
as it is also seen in the equivalent plot of fluxes (Fig.~\ref{broadha}b).

\begin{figure*}
\resizebox{\textwidth}{!}{
  \includegraphics{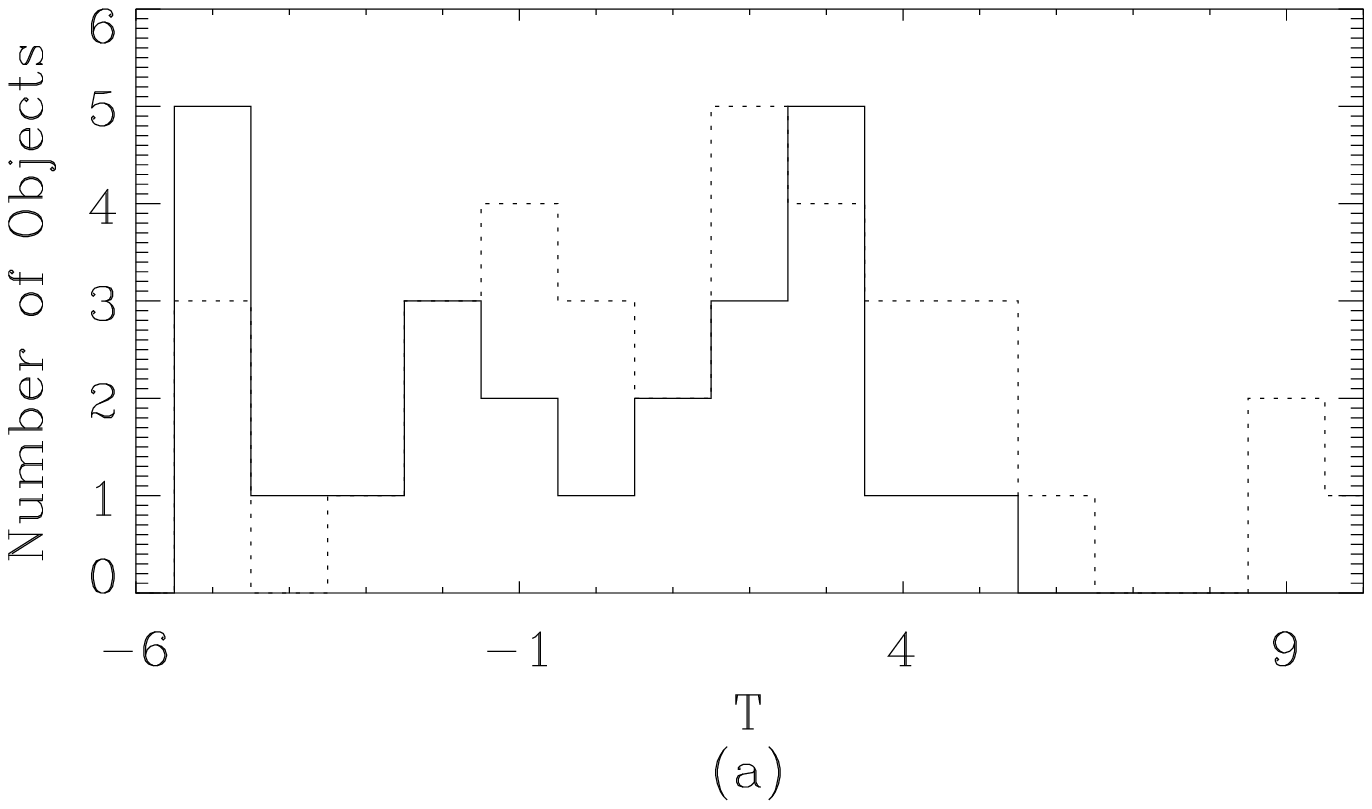}
  \includegraphics{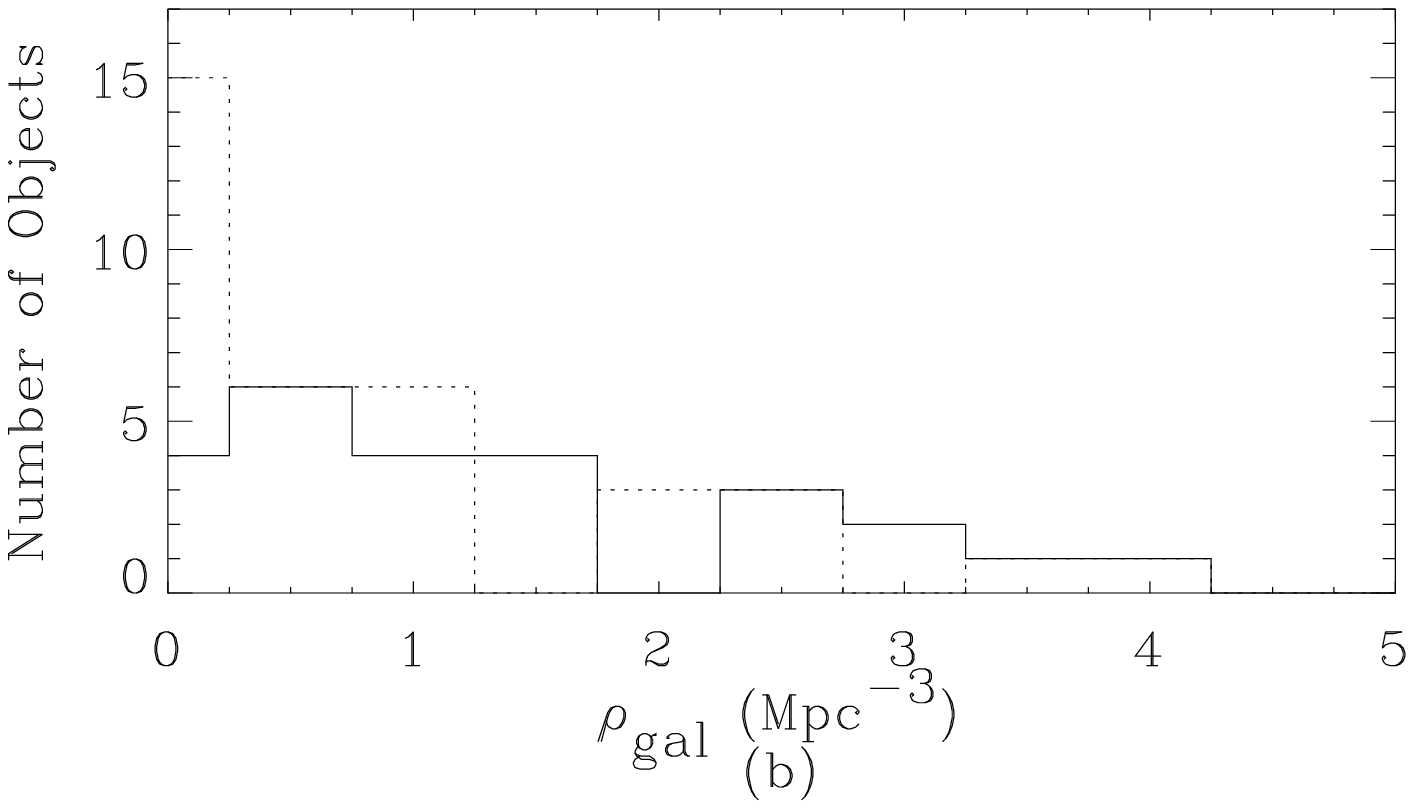}}
\caption{
 Histograms of 
 \textbf{(a)}~the galaxy morphological type, and 
 \textbf{(b)}~the local galaxy density, 
 comparing the \d19 LLAGNs detected at 2~cm (solid lines) and those not detected 
 at 2~cm (dotted lines). All values are as listed in H97a.}
\label{detvsnon}
\end{figure*}

\begin{figure*}
\resizebox{\textwidth}{!}{
  \includegraphics{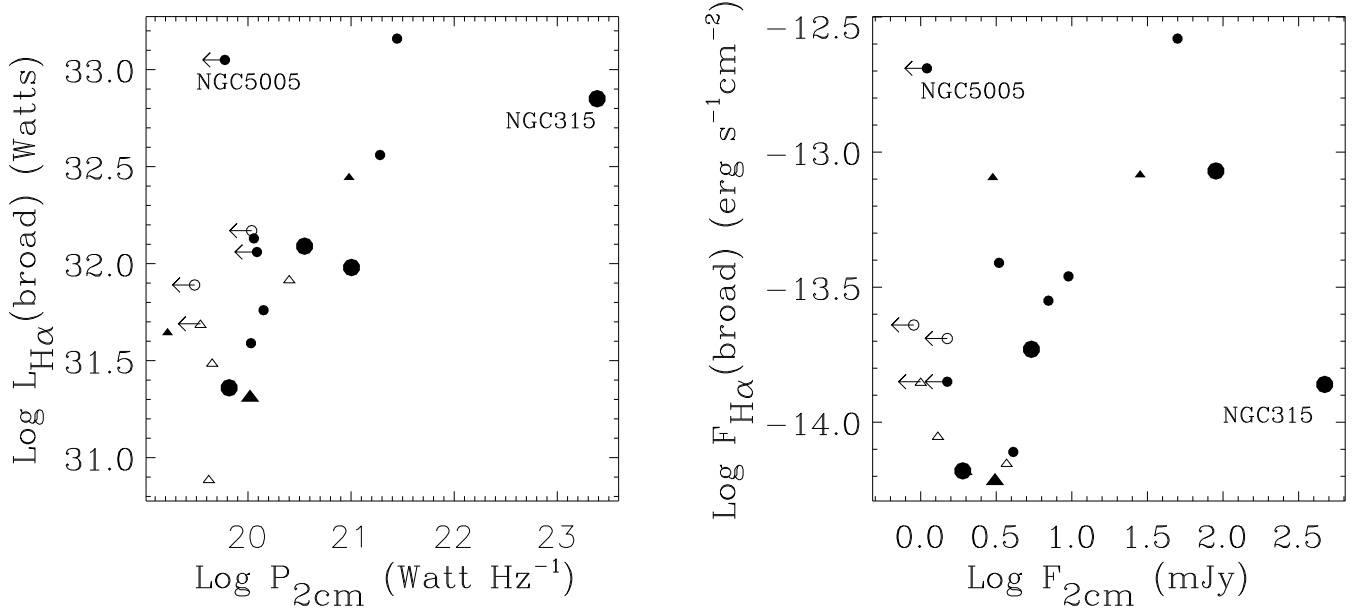}}
\caption{\textbf{(a)}~The luminosity of the broad component of \ha\ as a 
function of the 2~cm core (150~mas resolution) radio power for all (20) type~1.9 AGNs
in the Palomar sample for which photometric broad \ha\ fluxes are available.
LINERs are plotted as circles and Seyferts as
triangles. Nuclei with definite and probable detections of broad \ha\ \citep{hoet97c}
are shown with filled and open symbols, respectively. Elliptical nuclei are 
shown with large symbols; \textbf{(b)}~same as (a) but for fluxes instead of
luminosities.
}
\label{broadha}
\end{figure*}

\begin{figure*}
\resizebox{\textwidth}{!}{
  \includegraphics{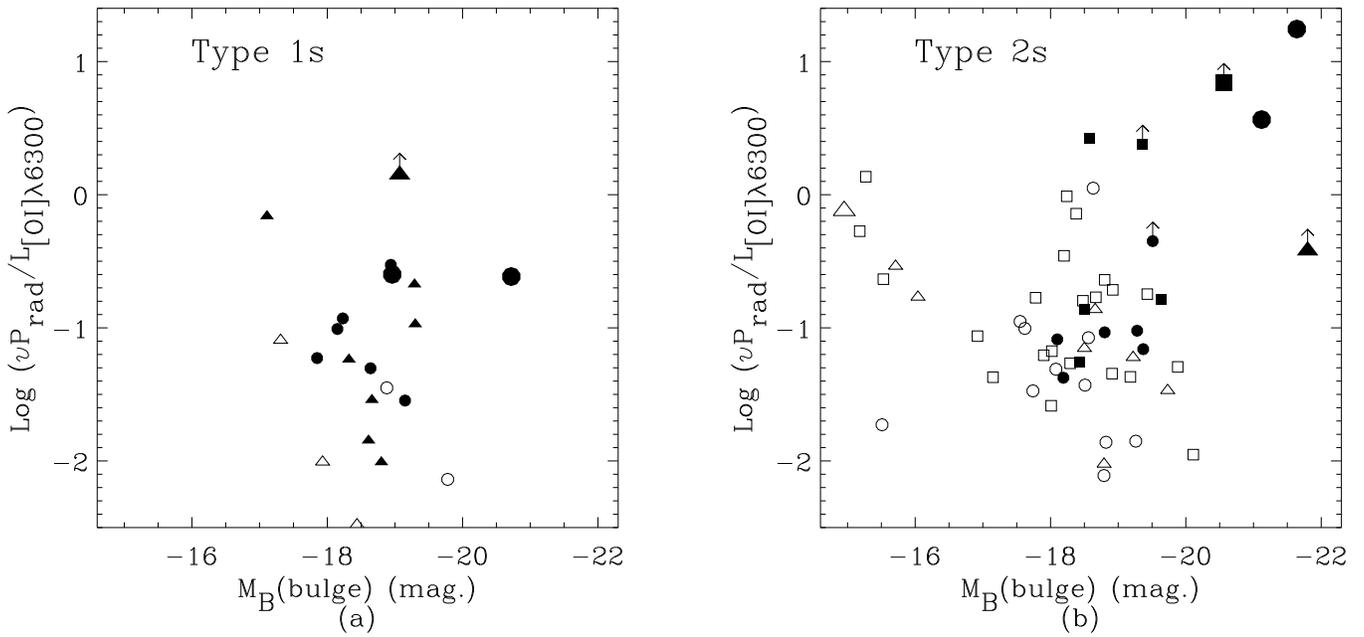}}
\caption{
 The ratio of 2~cm core (150~mas resolution) radio luminosity to 
 \fulloi\ luminosity as a 
 function of galaxy bulge magnitude in the blue band for type~1 LLAGNs 
 \textbf{(a)} and type~2 LLAGNs \textbf{(b)} in the distance-limited (\d19) sample. 
 LINERs are plotted as circles, Seyferts as triangles, and 
 transition nuclei as squares. Large symbols are used for elliptical galaxies
 and filled symbols are used for nuclei detected at 2~cm.
 Open symbols therefore represent upper limits to the true radio to 
 emission-line luminosity ratio. Nuclei with upper limits for both
 radio and \oi\ luminosities are not plotted.}
 \label{ratiorado1}
\end{figure*}

Other marginally significant differences between the radio-detected
and radio-non-detected nuclei include the following:
as compared to the non-detected LINERs and Seyferts, the 2~cm detected 
LINERs and Seyferts have higher [O~I] luminosities 
($\sim\,$93\% significance level; see Fig.~\ref{radvso1}), 
earlier morphological types (Fig.~\ref{detvsnon}a), higher bulge
luminosities (Fig.~\ref{ratiorado1}), and reside in regions with a higher galaxy density
(90\%--95\% significance level; Fig.~\ref{detvsnon}b).
The galaxy density, from \citet{tul88} and listed in H97a, is the 
density of all galaxies brighter than M$_{\rm B}$~= $-$16~mag 
in the vicinity of the object of interest.
In other words, 50\% of all 2~cm non-detected Seyferts and LINERs reside in 
galaxies which are located in regions of local galaxy density 
$<\!0.5\,$Mpc$^{-3}$, as compared to only 16\% for 2~cm detected Seyferts 
and LINERs. The difference in the galaxy density is primarily a result
of galaxies with more massive bulges being in regions of higher galaxy 
density:
the three most massive ellipticals in the sample (all radio-detected)
are at the centers of clusters ($\rho_{\rm gal}\,>\,3$) while the four 
galaxies with the least massive bulges are in regions with 
$\rho_{\rm gal}\,<\,1$.
When these very high and very low bulge mass galaxies are excluded, the 
difference in galaxy densities between the detected and non-detected
galaxies is no longer significant.
A comparison of the ratio of radio to \oi\ luminosity (Fig.~\ref{ratiorado1}) for
radio detected and non-detected LLAGNs reveals
that the ellipticals in the sample tend to have higher
L$_{\rm rad}$/\Loi\ ratios than the non-ellipticals for
both type~1 and type~2 nuclei. The highest ratios are in the most massive
type~2 ellipticals. 
Our radio detection limits are not low enough to reveal any clear difference
in this ratio between the radio detected and non-detected nuclei.

Most of the above factors are consistent with the idea that our radio 
observations have detected the more luminous LINERs and low-luminosity 
Seyferts, and that deeper radio observations of the sample will reveal a
higher incidence of compact radio cores. 

\subsection{Transition Nuclei} 
The only{\footnote{NGC~1161, tentatively identified as a transition nucleus 
(H97a) and listed in H97c as having a definite broad \ha\ line is not in 
the Palomar sample \citep{hoet95}.}} transition nucleus (out of 64 in 
the Palomar sample) which may have detected broad \ha\ emission is NGC~2985 
(at D\,= 22.4~Mpc); this detection is doubtful (H97c).
Thus it is most appropriate to compare transition nuclei to other type~2 
LLAGNs in the sample. Although the detection rate of Seyferts and LINERs 
of both types is much higher than that of transition nuclei, the detection 
rate for type~2 LINERs and type~2 Seyferts (10/38 or 26\%) is similar to that 
for all transition nuclei (6/35 or $\sim\,$17\%). Asurv tests indicate that 
the 2~cm radio powers and the [O~I]~$\lambda$6300 luminosities
of the transition nuclei are not significantly different from those of the
type~2 LINERs and type~2 Seyferts.
One way to test for a difference between the central engines of transition
nuclei and other type~2 LLAGNs is to look for a systematic difference in the
ratio of radio to \fulloi\ luminosities, since \fulloi\ is expected to
trace the `LINER' component rather than the star formation component. 
Our data do not show any clear difference (Fig.~\ref{ratiorado1}), but in
view of the large number of upper limits to $\nu$P$_{\rm rad}$/\fulloi\
deeper radio imaging is required to address this issue. 
Not surprisingly (Sect.~\ref{seclinllsey}), if we 
compare the transition nuclei to both type~1 and type~2 LINERs and Seyferts, 
then Asurv tests indicate that the 2~cm radio powers of the transition nuclei 
are lower at the 99.5\% significance level. 
The \fulloi\ luminosities of the transition nuclei are also
lower than those of the Seyferts and LINERs in the sample at the 97\%--99\% 
significance level.

\begin{figure}
\includegraphics{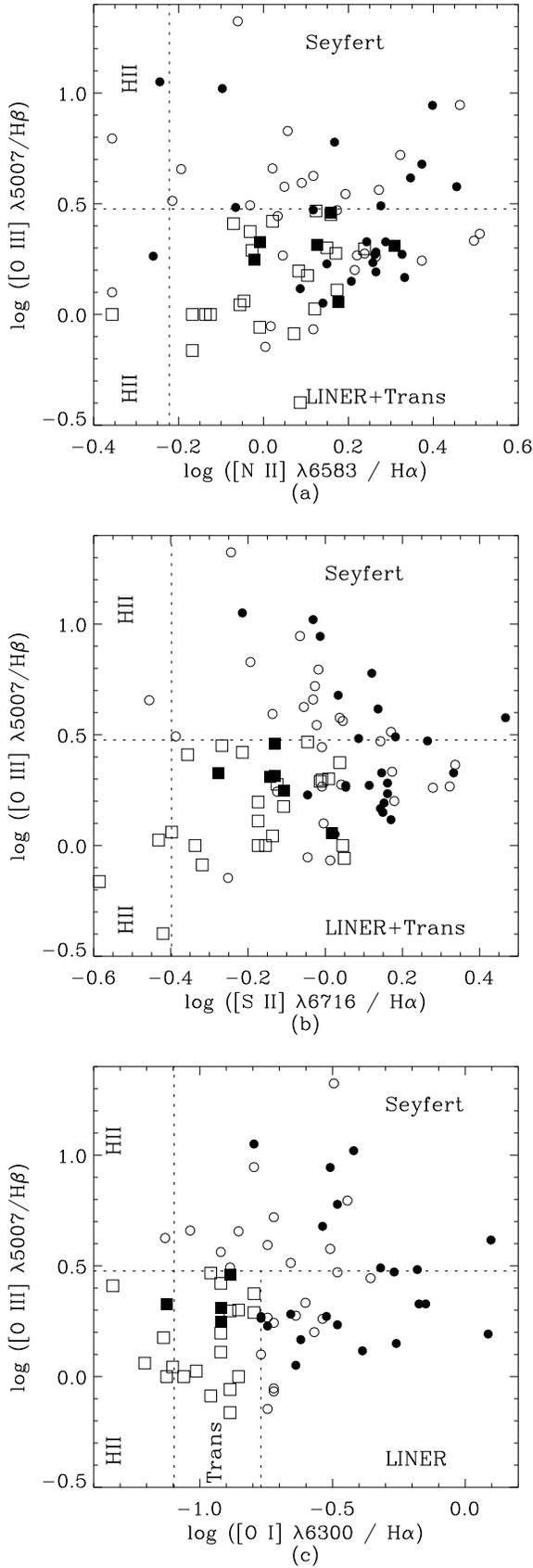}
\caption{
 Plots of standard diagnostic emission-line ratios for the \d19 LLAGNs. For 
 clarity, nuclei with upper limits to any one of the four emission-lines in 
 each respective panel are not plotted.
 Emission-line fluxes have been taken from H97a. LINERs and low-luminosity 
 Seyferts are plotted as circles and transition nuclei are plotted as 
 squares. Filled symbols are used for nuclei detected at 2~cm, and open 
 symbols for nuclei not detected at 2~cm. The dotted lines show the 
 spectral classification cutoffs used by H97a.}
\label{diagnostic}
\end{figure}

Are there any clear differences between the 6 detected and the 29 
non-detected transition nuclei? 
NGC~4552 is a massive elliptical and NGC~4419 and NGC~5866 have ambiguous 
identifications of broad lines (H97c)  which may (see 
Sect.~\ref{seclinllsey}) be related to their radio core detections. Another 
likely factor is differences in the emission-line ratios 
(Fig.~\ref{diagnostic}). The ratios plotted in Fig.~\ref{diagnostic} are 
used \citep[e.g.][]{veiost87} to distinguish between Seyferts, LINERs and 
\hii-region type nuclei. 
The dotted lines show the spectral classification cutoffs used by H97a.
Clearly the transition nuclei detected at 2~cm (filled squares) are
preferentially found among transition nuclei which lie close to the
LINERs and Seyferts in the plots (while the \nii/\ha\ and \sii/\ha\ ratios 
were not used for spectral classification it is clear that transition 
nuclei do not attain as high values of these ratios as LINERs and Seyferts).
Two of the detected transition nuclei, NGC~3627 and NGC~3628, are 
an interacting pair, with the activity classification of NGC~3627
bordering on a Seyfert 2.0 (H97a).  The classification of the elliptical 
NGC~4552 as a transition nucleus is uncertain as its H$\beta$ flux is 
uncertain at the 30\%--50\% level (H97a).
The median morphological type is T$^{\rm median}$~=~3 for the undetected
transition nuclei, while the detected transition nuclei have T values of
$-$5, $-$1, 1, 3, 3 and 3.
Finally, there are no significant differences in the \oi\ luminosities
and the local galaxy density between the detected and non-detected 
transitions.
Therefore, it appears that the strongest factor which determines the
probability of detecting a radio core in a non-elliptical transition nucleus 
is the proximity of its emission-line ratios to ``pure'' LINERs and 
Seyferts.  This gives support to the view that at least some transition 
nuclei have both ``pure'' LINER/Seyfert and \hii\ region components, with the 
detectability of the radio core determined by the strength of the former 
component. 
\citet{filet00} have detected arcsec-scale, 3.6~cm cores  with
a flat radio spectrum, in 5 of 25 transition nuclei from the Palomar
spectroscopic sample (two of their five cores, NGC~3627 and NGC~4552,
are also found to be compact and flat spectrum by us). They find that the
detected transition nuclei are preferentially in early-type galaxies; our
results are consistent with this.

\subsection{\hii\ Nuclei}

While we have not investigated \hii\ nuclei from the Palomar sample in the
radio, it is instructive to check in what ways galaxies with \hii\ region nuclei 
may differ from those with LLAGNs.
As shown by \citet{hoet97b}, the \hii\ nuclei in  the Palomar sample
are in later-type hosts as compared to the LLAGNs. This is also
true for the subset of the Palomar sample at \d19.
Further, the median absolute magnitude in the B-band of the bulge for the
\hii\ region galaxies at \d19
(M$_{\rm B}^{\rm median}$(bulge)~= $-$16.83) is
significantly fainter than the equivalent median value for the low-luminosity
Seyferts and LINERs at \d19 (M$_{\rm B}^{\rm median}$(bulge)~=
$-$18.65). The median values have been obtained from the values of
M$_{\rm B}$(bulge) listed in H97a.
The differences between the \d19\ galaxies in the Palomar sample with \hii\ nuclei
and those with LLAGNs are thus deeper than just their nuclear spectral
classification. 
Thus, while the \hii\ region nuclei may also harbor accreting black holes, 
these will form a different population from those in LINERs, low-luminosity 
Seyferts, and transition nuclei.

\section{The Influence of the Galaxy Bulge on the Properties of the 
         Distance-Limited Sample}
\label{secgalbulge}
The bulge luminosity of the host galaxy 
is known to strongly influence the probability of finding a nuclear
radio source in radio galaxies \citep{auret77}, and the power of the 
nuclear radio emission in early-type galaxies \citep{sadet89}.
Further, for radio and Seyfert galaxies, the nuclear centimeter radio 
luminosity at arcsec-scales is correlated with the bulge luminosity over 
six orders of radio power \citep{nelwhi96}. The bulge luminosity is also 
known to affect the nuclear emission-line luminosity in Seyfert galaxies 
\citep{whi92b}, radio-galaxies \citep{zirbau95},
and even ``normal'' early-type galaxies \citep{sadet89}.
In this section, we test whether the above relationships hold at the lower 
radio and emission-line luminosities of our sample of LLAGNs.

\subsection{Emission-Line Luminosity}
\label{secbulemi}

\begin{figure*}
\resizebox{\textwidth}{!}{
\includegraphics[bb=72 500 540 720,width=\textwidth,clip]{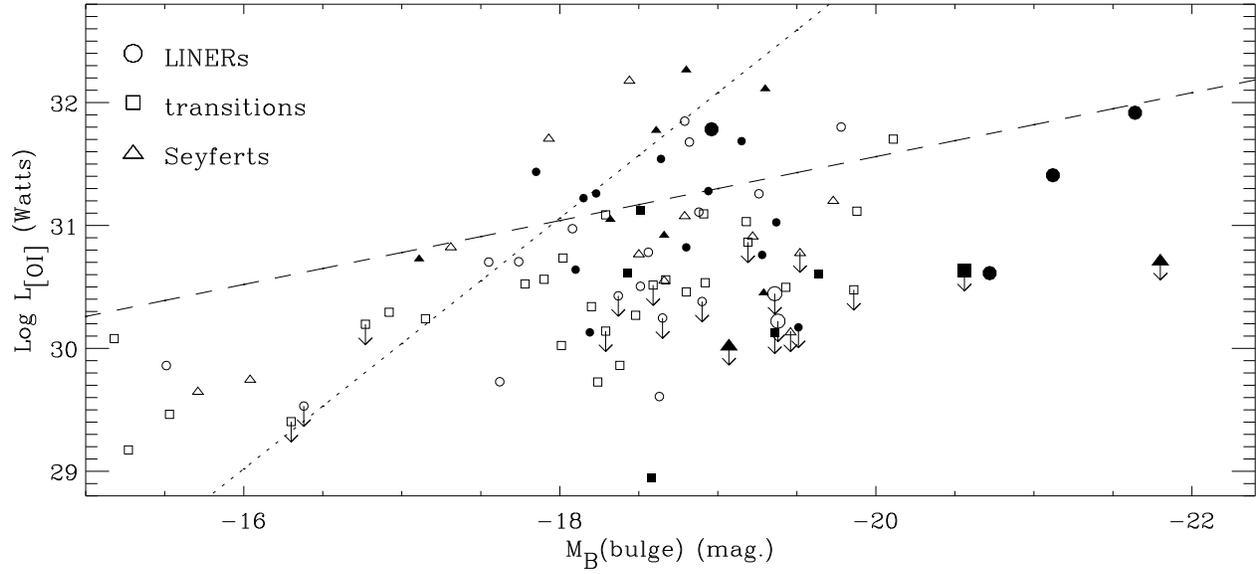}}
\caption{
 The dependence of the luminosity of the \fulloi\ emission-line on the absolute 
 magnitude in the blue band of the galaxy bulge for the distance-limited
 (\d19) sample.
 LINERs are plotted as circles, Seyferts as triangles and
 transition nuclei as squares. Large symbols are used for 
 elliptical galaxies and filled symbols for radio detected nuclei.
 Also shown is the low-luminosity extrapolation of the linear fit to the same
 relationship for the luminous Seyferts in the \citet{whi92a} compilation (dotted line; 
 see text) and an indicative low-luminosity extrapolation of the linear fit to the
 same relationship for FR~I radio galaxies from \citet[][dashed line; see text]{zirbau95}.}
\label{bulo1}
\end{figure*}

Among the radio detected nuclei, the \fulloi\ luminosity is not clearly 
related to the galaxy bulge magnitude (Fig.~\ref{bulo1}). 
Also, for $-17\,>$ M$_{\rm B}$(bulge) $>\,-20$ the radio detected nuclei 
and non-detected nuclei are fairly well mixed at any given bulge magnitude.
It is only at the lowest and highest bulge magnitudes that all nuclei are 
undetected and detected in the radio, respectively.
Also shown in the figure are the low-luminosity extrapolations
of the relationship between \Loi\ and M$_{\rm B}$(bulge)
for `classical' Seyferts (dotted line; for these nuclei
the \fulloiii\ luminosity from \citet{whi92a} was converted to 
the \oi\ luminosity using a factor of 0.04, typical of Seyferts 
\citep{veiost87}) and an indicative line for FR~I radio galaxies (dashed 
line; the total \fullhanii\ luminosity from \citet{zirbau95} was converted to an \oi\
luminosity using a factor of 0.08, typical of the LINERs in our sample).
\citet{zirbau95} list total emission-line luminosities for radio galaxies; typically 
more than half of this luminosity is from the nuclear ($<\,$2.5~kpc) region 
\citep{bauhec89}.
It is notable that, while several Seyfert galaxies fall close to the low-luminosity
extrapolation of `classical' Seyferts, many lie much further away,
the farthest being NGC~4168, NGC~4472 (both ellipticals),
NGC~4565, and NGC~4725. The elliptical LINER with a high \oi\ luminosity for its 
bulge magnitude is NGC~4278.
The radio-detected Seyferts are indistinguishable from the 
other radio-detected LLAGNs and, when considered together, the 
radio-detected LLAGNs are closer to the low-luminosity extrapolation of 
FR~I radio galaxies than to that of `classical' Seyferts.

\begin{figure*}
\resizebox{16cm}{!}{   
\includegraphics{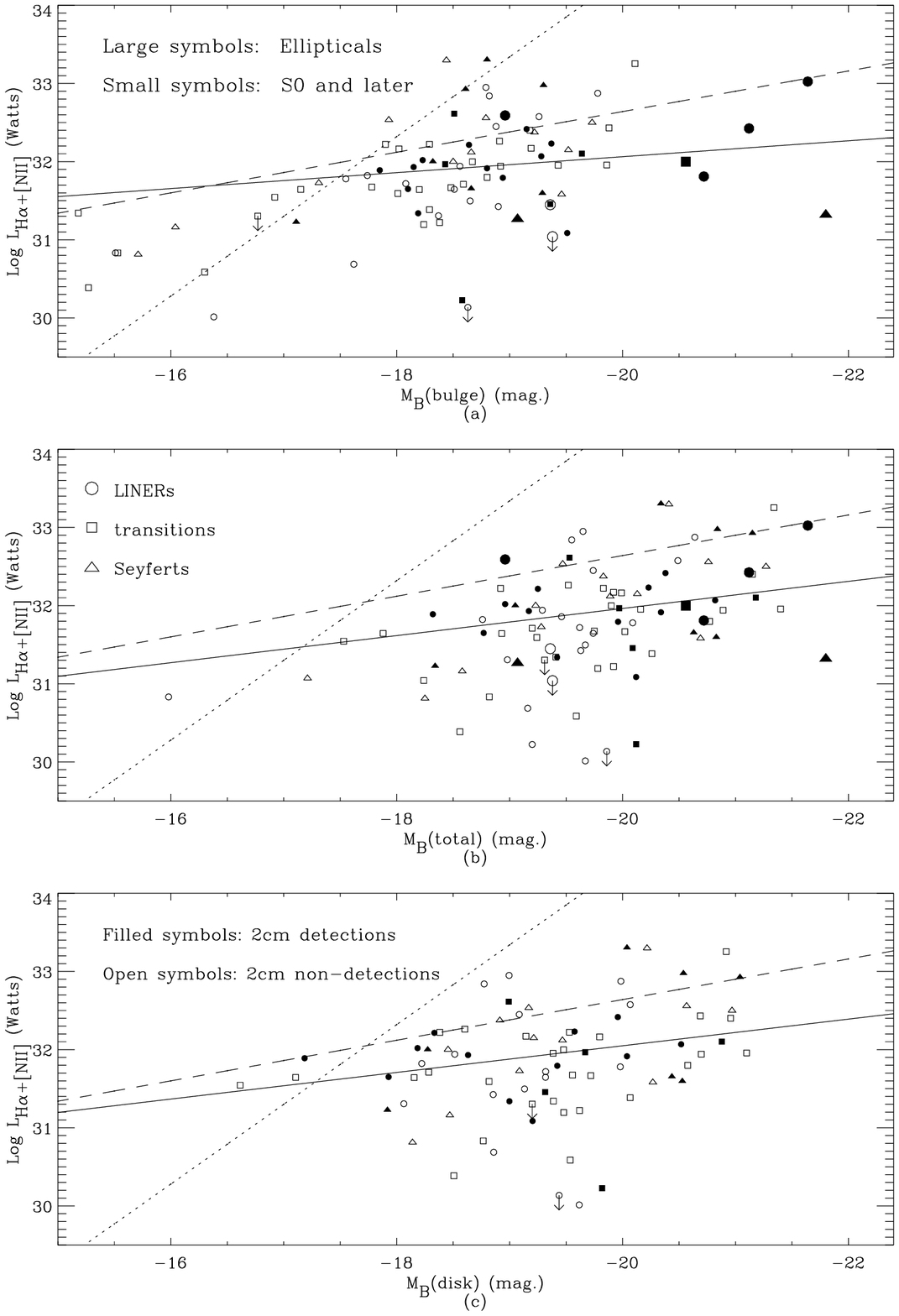}}
\caption{
 The dependence of the luminosity of the \fullhanii\ emission-lines on the 
 absolute magnitude in the blue band of 
 \textbf{(a)}~the galaxy bulge;
 \textbf{(b)}~the entire galaxy; and
 \textbf{(c)}~the galaxy disk, for the distance-limited (\d19) sample.
 Some of the galaxies fall beyond the left and bottom boundaries of the plots.
 Symbols are as in the previous figure.
 Linear fits to the radio-detected nuclei are shown as the solid line in
 each panel.
 Also shown in each panel is the low-luminosity extrapolation of a linear fit to the
 \Lhanii\ vs M$_{\rm B}$(bulge) relationship for 
 luminous Seyferts in the Whittle (1992) compilation (dotted line; see text)
 and FR~I radio galaxies from \citet[][dashed line]{zirbau95}.}
\label{bulemi}
\end{figure*}

Since the \ha\ luminosity is often used to represent the emission-line
luminosity for AGNs we now look at the dependence of this quantity on the
galaxy luminosity (Fig.~\ref{bulemi}).
The \hanii\ luminosity is significantly correlated with galaxy total 
luminosity separately for all ellipticals and all non-ellipticals in 
the Palomar sample (Fig.~\ref{bulemi}b; Table~\ref{tabstatll96}).
Figure~\ref{bulemi}a suggests that, at a given galaxy bulge magnitude, the 
radio detected ellipticals (except for NGC~4278) have lower emission-line 
luminosities than the radio detected non-ellipticals, though the number 
of ellipticals is only 10.
However, when the total galaxy luminosity is considered 
(Fig.~\ref{bulemi}b), the ellipticals and non-ellipticals detected in
the radio fall more or less together, i.e.  the galaxy total luminosity, 
rather than the galaxy bulge luminosity, appears to be a better predictor 
of the nuclear emission-line luminosity among radio detected LLAGNs. 
The low numerical significance of the statistical tests involving the 
radio-detected galaxies (Table~\ref{tabstatll96}) may simply reflect
the small number of nuclei considered; the correlation between 
emission-line luminosity and galaxy bulge or total luminosity becomes 
significant when we also consider 
the 2~cm non-detections at \d19 (Table~\ref{tabstatll96}, bottom).
A linear fit to all radio-detected nuclei in Fig.~\ref{bulemi} gives
\Lhanii$\propto$~L$_{\rm B}^{0.43}$(total) and
$\propto$~L$_{\rm B}^{0.26}$(bulge).

As with the \oi\ luminosity, the \hanii\ luminosities of the LLAGNs fall 
closer to the low-luminosity extrapolation of FR~I radio galaxies 
\citep{zirbau95} than to that of `classical' Seyferts (Fig.~\ref{bulemi}). 
The \hanii\ luminosities of the `classical' Seyferts were
derived by converting the \fulloiii\ luminosities listed in
\citet{whi92a} using standard emission-line ratios in Seyferts:
\ha/H$\beta\,=\,$3.1, \fulloiii/H$\beta\,=\,$10, and
\fullnii/\ha$\,=\,$1 \citep{veiost87}.
As for the plot of \Loi\ vs M$_{\rm B}$(bulge) given in Fig.~\ref{bulo1},
Fig.~\ref{bulemi} shows that for a given galaxy total or bulge luminosity, 
the radio detected and radio non-detected galaxies have similar nuclear
emission-line luminosities.  It is only at the lowest and highest bulge luminosities 
that all galaxies are undetected and detected, respectively, in our radio survey.

\subsection{Core Radio Power}
\label{secbulrad}

\begin{figure*}
\resizebox{17cm}{!}{
\includegraphics{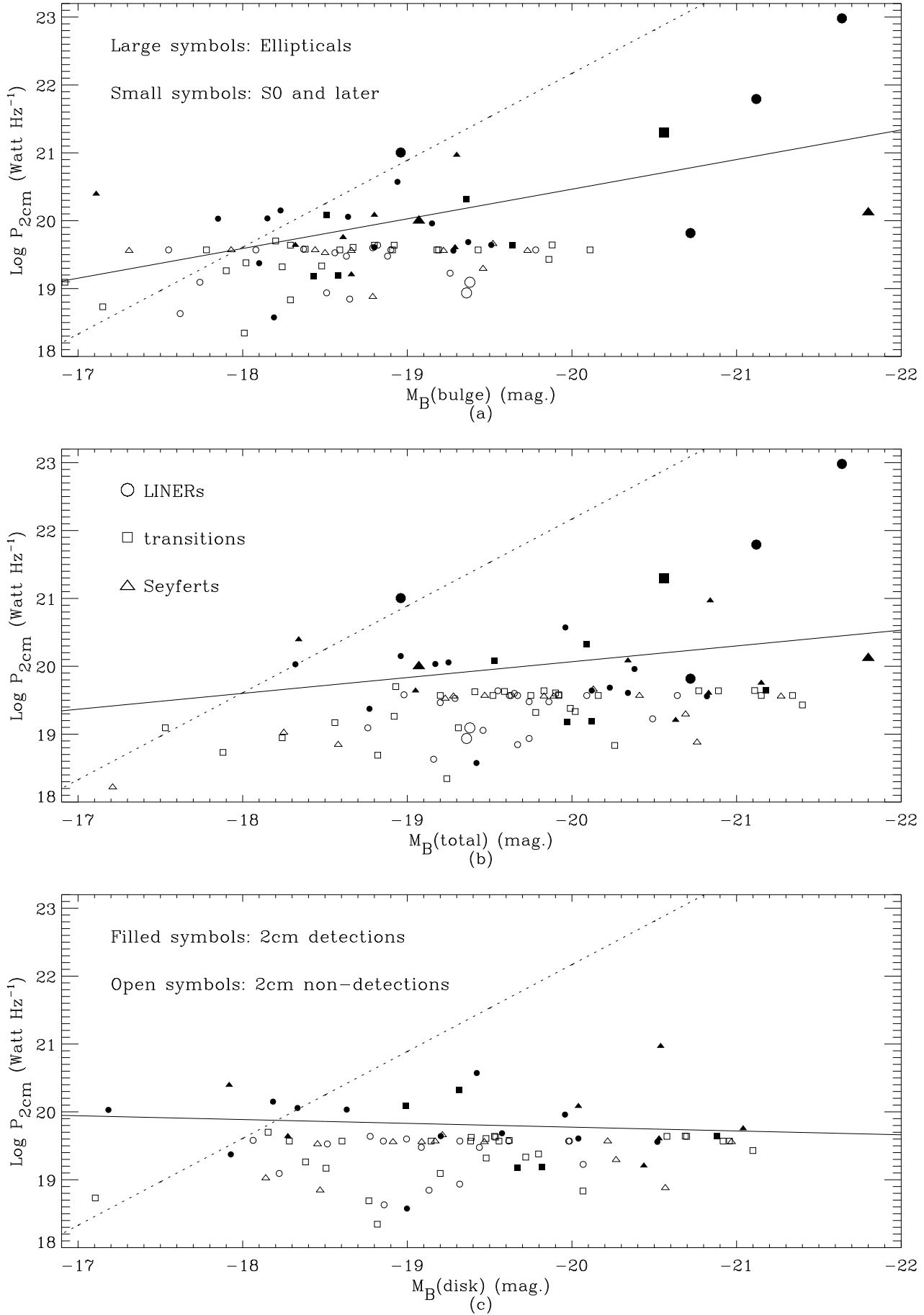}}
\caption{
 As for Fig.\ref{bulemi}, but with the 2~cm core (150~mas resolution) radio 
 power as the $y$ axis.
 Some of the galaxies fall beyond the left and bottom boundaries of the plots.
 Each panel also shows the best linear fit to the radio detected nuclei
 (solid line), and the low-luminosity extrapolation of a linear fit to the core 
 radio power vs bulge luminosity relationship in both luminous Seyferts and 
 (FR~I and FR~II) radio galaxies \citep[][dotted line]{nelwhi96}.}
\label{bulrad}
\end{figure*}

The radio power of the radio-detected nuclei which are not in ellipticals
does not depend on the galaxy total, bulge, or disk luminosity; linear fits 
in all cases give a line with approximately zero slope (Fig.~\ref{bulrad}). 
The upper limits to the radio emission of the undetected (small open symbols 
in Fig.~\ref{bulrad}) non-ellipticals strongly support this result.
The few elliptical nuclei in the sample show a variation of 1 to 3 orders 
of magnitude in radio power for a given galaxy luminosity.
The ellipticals in the sample generally have higher radio 
powers and higher galaxy luminosities than the non-ellipticals in the sample.
Probably as a result of this, when all radio detected nuclei in the sample 
(i.e. both ellipticals and non-ellipticals) are considered together, the 
radio power is found to be correlated with the galaxy bulge luminosity 
(but not the galaxy total luminosity; Table~\ref{tabstatll96}), with linear 
fit P$^{\rm core}_{\rm 2cm}~\propto$ L$_{\rm B}^{1.1}$(bulge).
The slope of this dependence is much shallower than that seen over 6 orders 
of magnitude of core radio power by Nelson \& Whittle (1996) for both 
powerful radio (FR~I and FR~II) galaxies and Seyfert (type~1 and type~2) 
galaxies (dotted line in Fig.~\ref{bulrad}a; 
P$^{\rm core}_{\rm 2cm}~\propto$ L$_{\rm B}^{3.2}$(bulge)). 
It is also shallower than the relationship 
displayed by ``normal'' nearby elliptical and S0 
galaxies{\footnote{We used the relationships derived by \citet{nelwhi96}
and \citet{sadet89} after changing $H_0$ to the value used 
in this paper and transforming their 20~cm luminosities to 2~cm assuming
optically-thin synchrotron emission (spectral index $-$0.7). Both these 
transformations change the intercept of the line but not the slope.}} 
\citep[the 30th percentile of the radio power rises as
L$_{\rm B}^{2.2}$;][]{sadet89}.
Figure~\ref{bulrad} shows that the core radio powers of LLAGNs
are less than what would be expected from the low-luminosity 
extrapolations of more powerful AGNs.  Factors which could affect
the slopes and radio powers include the differences in the
frequencies and resolutions between the various datasets. Given the longer 
wavelength (20~cm) and lower resolution ($1{\arcsec}\,-\,3{\arcsec}$) of the 
data used by \citet{nelwhi96} and \citet{sadet89}, the nuclear radio emission 
is more likely to include emission from arcsec-scale radio jets and 
star-formation activity in the bulge in their data
\citep[see e.g. Sect. 11.2 of][]{whi92b} as compared to our higher
resolution 2~cm data.
High resolution radio observations, capable of isolating the flat spectrum
pc-scale cores in the more powerful AGNs, are needed for comparison
with our data.

\subsection{The Interplay Between Core Radio Power and Emission-Line 
         Luminosity}
\label{secinterplay}
The interplay between radio and emission-line luminosity can be better
understood in the context of the following two factors (discussed above):
(1)~for all LLAGNs considered together, the 2~cm core power 
(P$^{\rm core}_{\rm 2cm}~\propto$ L$_{\rm B}^{1.1}$(bulge))
increases more rapidly than the nuclear emission-line luminosity 
(\Lhanii$\propto$~L$_{\rm B}^{0.26}$(bulge))
with increasing bulge luminosity, as is also the case for ``normal'' 
elliptical and S0 galaxies \citep{sadet89};
(2)~with increasing disk luminosity, the emission-line luminosity increases,
but the radio power does not.

\begin{figure*}
\sidecaption
\includegraphics[width=13cm]{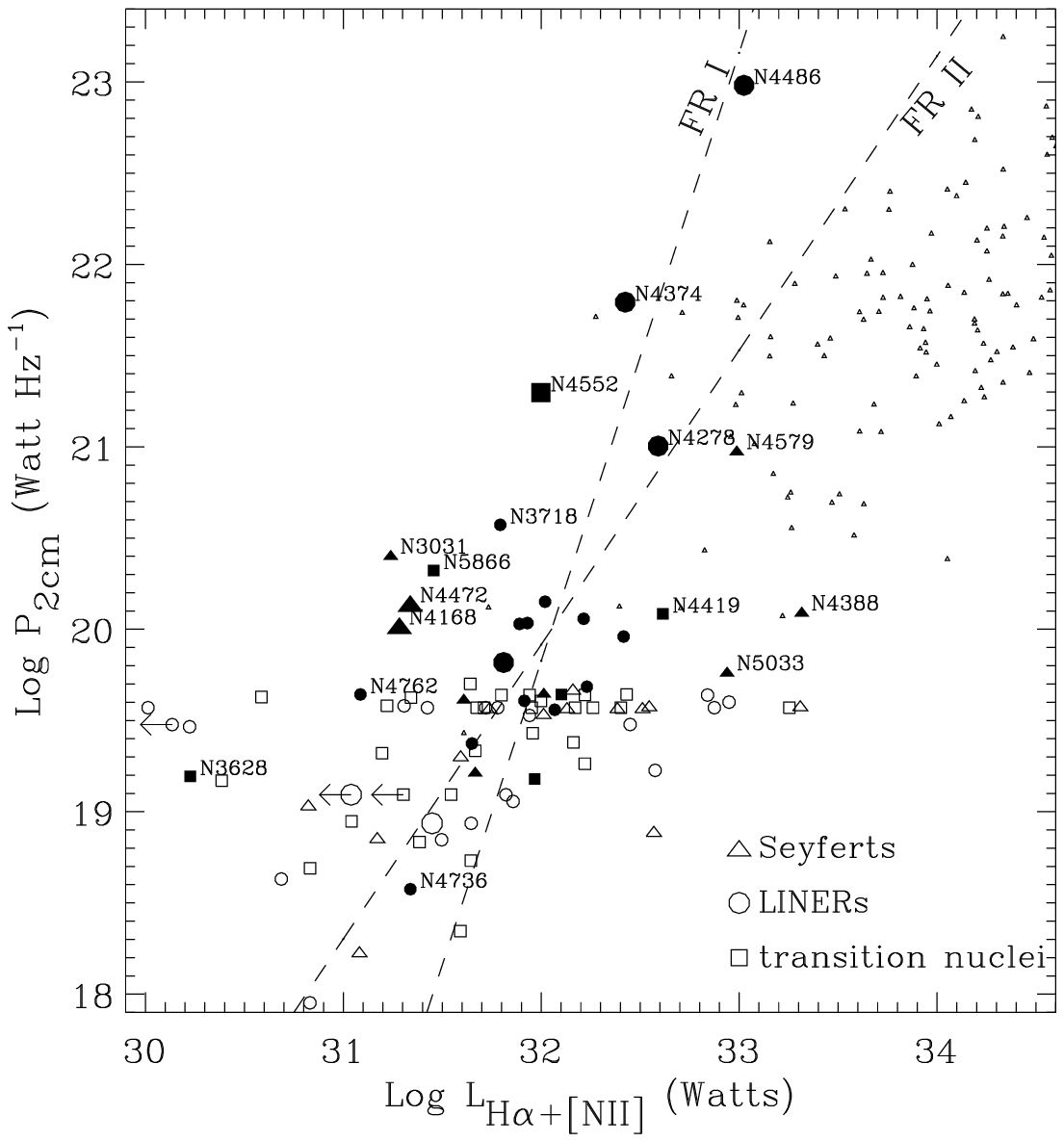}
\caption{
 A plot of 2~cm core (150~mas resolution) radio power versus 
 nuclear \fullhanii\ 
 luminosity for all LLAGNs in the distance-limited (\d19) sample.
 Symbols are as used in Fig.\ref{bulo1}.
 ``Classical'' Seyfert galaxies (from Whittle 1992a) are plotted as the smallest
 triangles (for these the \oiii\ luminosity was converted to an \hanii\ 
 luminosity using standard flux ratios for Seyferts, see text).
 Also shown are the low-luminosity extrapolations of linear fits to the same
 relationship for FR~I and FR~II radio galaxies (dashed lines; Zirbel \& Baum 1995,
 see text).}
\label{radvshan2}
\end{figure*}

A plot of core radio power against \fullhanii\ luminosity for all LLAGNs at 
\d19 is shown in Fig.~\ref{radvshan2}. 
This type of plot has been used to separate 
`radio-loud' AGNs from `radio-quiet' ones \citep[e.g.][]{xuet99}.
Also plotted in the figure are the low-luminosity extrapolations
of the same relationship for FR~I and FR~II radio galaxies from 
\citet[][dashed lines]{zirbau95}, and the sample of `classical' Seyferts from 
\citet[][small triangles]{whi92a}. The values quoted in these papers have
been converted to a Hubble constant of 75 km s$^{-1}$ Mpc$^{-1}$, and the 
Seyfert \hanii\ luminosities have been derived from the \fulloiii\ 
luminosity as described in Sect.~\ref{secbulemi}.
Though there is considerable scatter, factor (1) above causes a
general trend for high bulge-mass LLAGNs to occupy the top center of the
plot and low bulge-mass LLAGNs to occupy the lower left. However, factor 
(2) allows many of the galaxies with low bulge luminosities to sit to the 
right of their compatriots with higher bulge luminosities by dint 
of their higher disk luminosities.

It is very interesting that most of the radio-detected LLAGNs fall in the 
region of Fig.~\ref{radvshan2} where the low-luminosity extrapolations of 
the radio-loud (FR~I and FR~II radio-galaxies) and radio-quiet 
\citep[Seyfert galaxies from the compilation of][]{whi92a} nuclei intersect. 
There is thus no evidence for a radio-loud - radio-quiet divide in the 
LLAGNs.
Most of the radio detected LINERs and transition nuclei are found close to 
the low-luminosity extrapolations of the FR~I and FR~II galaxies.
Some of the LLAGNs are offset to higher emission-line luminosities from
the FR~I and FR~II relationships. Essentially all of these LLAGNs are in
disk-type galaxies, with disk gas probably contributing to the emission-line 
luminosity. The operational distinction between `classical' Seyferts and 
low-luminosity Seyferts is log\,[(\Lhanii)/W]\,$\sim$ 33, so 
it is not surprising that low-luminosity Seyferts near this luminosity fall 
near the low luminosity extrapolation of `classical' Seyferts.
It is notable in Fig.~\ref{radvshan2} that the elliptical galaxies tend to
have a higher radio power at a given emission-line luminosity than the S0
or disk galaxies, as we already inferred from Fig.~\ref{ratiorado1}.

\section{Radio Luminosity and Black Hole Mass}
\label{secoverallmdo}
\begin{figure*}
\resizebox{\textwidth}{!}{
\includegraphics{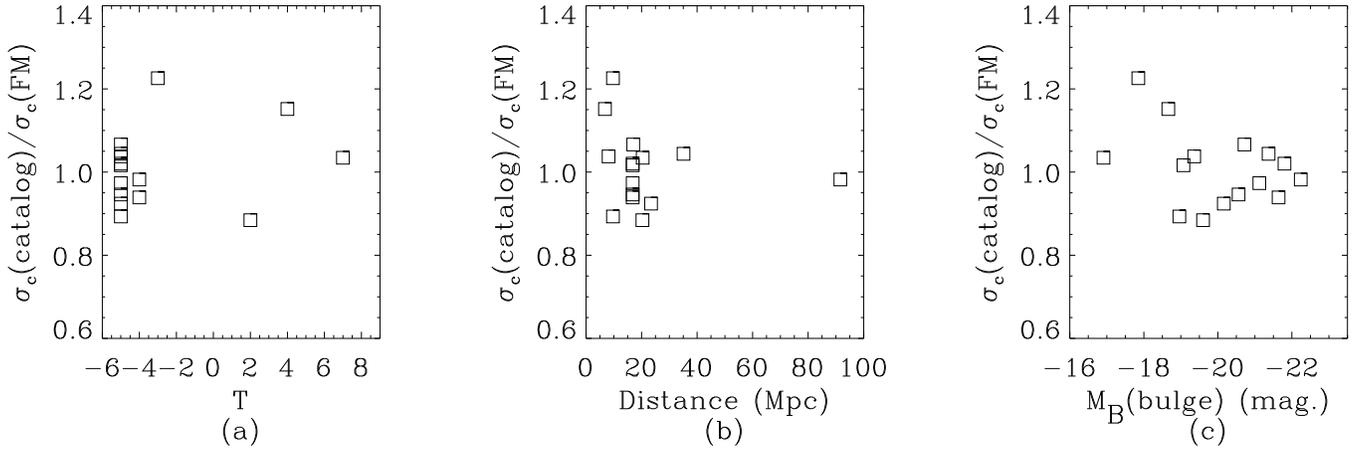}}
\caption{
 The ratio of the central velocity dispersion listed in either LEDA or Hypercat 
 to that listed in \citet{fermer00} and \citet{merfer01} is plotted against the 
 \textbf{(a)}~morphological type of the galaxy; 
 \textbf{(b)}~distance to the galaxy; and
 \textbf{(c)}~absolute magnitude in the B band of the galaxy bulge,
 for galaxies in the Palomar sample.
 The values from these sources agree to better than $\pm$20\%.}
\label{sigmachk}
\end{figure*}

We now investigate the relationship between the nuclear radio luminosity 
and the mass of the nuclear black hole in nearby galaxies. When a direct 
measurement of the black hole mass from high spatial-resolution kinematic 
measurements \citep{ricet98,gebet00} is not available, we use the relationship 
found by \citet{merfer01} between the black hole mass and the central 
stellar velocity dispersion, $\sigma_c$. 
\citet{merfer01} and \citet[][who used $\sigma_e$, the luminosity-weighted 
line-of-sight dispersion within the half-light radius]{gebet00} have
shown that this relationship is very tight with its accuracy currently
limited only by the measurement errors in $\sigma_c$ and \mmdo. Values of 
$\sigma_c$ have been taken, in increasing order of preference, from 
the Lyon-Meudon Extragalactic Database \citep[LEDA;][]{patet97},
\citet{pruet98} and its update in the `Hypercat' database{\footnote{
http://www-obs.univ-lyon1.fr/hypercat/}},
\citet{fermer00} and \citet{merfer01}.
\citet{fermer00} and \citet{merfer01} list the value of $\sigma_c$ after
correction for aperture size. They find that the aperture corrections for 
the galaxies in their study are of order a few km~s$^{-1}$ except for some 
more distant ($\gtrsim\,$60~Mpc) galaxies. Therefore, since most of the
galaxies we consider below are relatively nearby (typically $\leq\,$30~Mpc),
we have not made any aperture corrections to the values of $\sigma_c$ taken 
from other sources.
In support of this, a comparison between the values of $\sigma_c$ from
Ferrarese \& Merritt and those taken from LEDA or Hypercat agree to within 
20\% (Fig.~\ref{sigmachk}). There is no evidence that the difference in the
values of $\sigma_c$ increases with galaxy distance or for later morphological 
types. However, the difference between the two estimates may increase with 
decreasing galaxy bulge luminosity (Fig.~\ref{sigmachk}c).

In order to minimize confusion by radio emission from larger scale regions,
e.g. jets/lobes or star formation in the bulge, we use mainly sub-arcsecond
scale radio fluxes, and treat all radio flux measurements made at resolution 
$>\,150\,$mas as upper limits to the `core' radio flux. One may expect that 
even the $\leq$~150~mas-resolution data (typically $\leq$~15~pc) 
overestimates the flux from the central few hundred Schwarzschild radii. 
However, free-free and synchrotron self-absorption may be significant on
sub-pc-scales, so the central flux could be significantly attenuated
even at 2~cm \citep[e.g.][]{ulvet99}.
Most of the measurements are made at frequencies of 15~GHz or higher and
thus avoid some of the problems of contamination by radio outflows and
star formation related processes.
For the handful of measurements at frequencies lower than 15~GHz, we
converted to estimated 15~GHz fluxes using a spectral index of 0,
which is the median spectral index observed for mas-scale core radio
emission in LLAGNs \citep{naget01}. The lowest observed frequency we use is 
1.4~GHz (20~cm); thus using a spectral index in the range $\pm0.7$ instead of 0 
would change the extrapolated 15~GHz flux by a factor of $\leq$5.

\subsection{Results for the Distance-Limited Sample of LLAGNs}

\begin{figure*}
\sidecaption
\includegraphics[width=13cm]{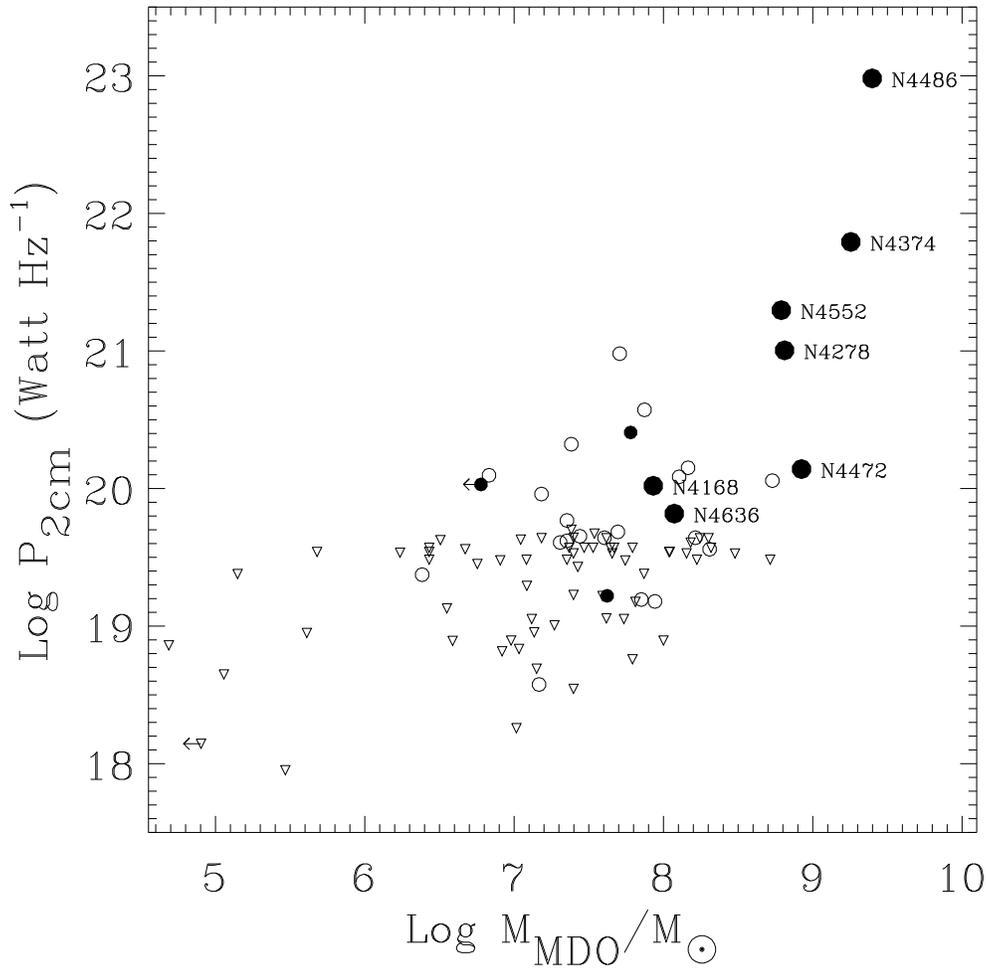}
\caption{
 Dependence of the core (150~mas resolution) 2~cm power on the mass of the 
 supermassive black hole for LLAGNs in the distance-limited (\d19) sample. 
 For the radio detections, large symbols are used for ellipticals (which are 
 labelled with their names) and filled symbols for the most reliable measurements 
 of MDO mass (see text). 
 All radio non-detections are shown by the downward pointing triangles. NGC~185
 lies beyond the bottom left corner of the plot.}
\label{llagnmdo}
\end{figure*}

The correlation between radio luminosity and bulge luminosity for LLAGNs has 
been explored in Sect.~\ref{secbulrad}. The basic result is that
the two quantities are correlated if all radio detections are considered
together, but not when only non-elliptical galaxies are considered.
The plot of radio luminosity vs. MDO mass (Fig.~\ref{llagnmdo}) also 
shows a strong correlation but considerable scatter. 
In fact the five 
massive ellipticals in the figure~- NGC~4278, NGC~4374, NGC~4472, NGC~4486, 
and NGC~4552~- have very similar MDO masses ($\sim\,$10$^9$~\Msun) but 
widely different radio powers, even at mas resolution. 
This large spread does not necessarily imply a large spread in the radio 
luminosities within a few hundred R$_s$ as there is evidence that the radio 
emission in all five of these sources is jet dominated \citep[see][]{naget01}.

\subsection{Results for a Larger Sample of Nearby Galaxies}
\label{secmdo}

We now consider the interplay between radio power, MDO mass and 
bulge luminosity in the larger sample (Sect.~\ref{secmdosample})
of relatively nearby galaxies. The core radio fluxes for the Seyferts and 
LINERs at D$\,>\,19\,$Mpc in the Palomar sample were taken from
\citet{naget00} and Nagar et al. (in preparation). These are 150~mas
resolution, 2~cm observations, like the observations of the \d19 sample
of LLAGNs. The core radio fluxes of the additional galaxies were taken from 
\citet{caret90},
\citet{conet90}, 
\citet{cra79}, 
\citet{craet93}, 
\citet{dhaet98},
\citet{dimet01},
\citet{fabet89}, 
\citet{fergio87}, 
\citet{fraet98}, 
\citet{galet97}, 
\citet{gelfom84}, 
\citet{gioet98}, 
\citet{graet81}, 
\citet{jon74},
\citet{jonweh97}, 
\citet{kelet98}, 
\citet{kriet98}, 
\citet{lauet97}, 
\citet{naget99},
\citet{sadet95}, 
\citet{sleet94}, 
\citet{troet98},
\citet{venet93},
\citet{whiet97}, 
and \citet{wrohee91}. 
When kinematically-determined MDO masses were not available, we estimated
the MDO mass from $\sigma_c$, as outlined in the previous sections.
When available, distances have been taken from the same reference as that
providing the black hole mass. 
Galaxy morphological types, corrected total magnitudes in the B band, and
distances for the remaining galaxies have been taken from LEDA.
All quantities were converted to the value of $H_0$ --
75~km~s$^{-1}$ Mpc$^{-1}$ -- used in this paper.
In all, we found that 154 of these galaxies, 37 of which are ellipticals, have 
values for all three of the following: estimated MDO mass, core radio flux, and 
bulge luminosity. Sixty six of these 154 galaxies have a radio core 
detected at resolution better than 150~mas. 

The nearby galaxies, spanning 9 orders of magnitude in radio power, 6 orders 
of magnitude in MDO mass, and 5 orders of magnitude in bulge magnitude, are 
plotted in Fig.~\ref{mdorad}\footnote{NGC~4278 has moved significantly between 
Figs.~\ref{llagnmdo} and \ref{mdorad} as H97a use a distance of 9.7~Mpc 
while \citet{ricet98} use a distance of 17.5~Mpc, which is used in 
Fig.~\ref{mdorad}.}. With such a large range of values, it is clear
that all three quantities are correlated. Asurv tests on all 154 galaxies
and on only the 66 with radio cores at $\leq$~150~mas resolution, indicate 
that the radio power is correlated with both the MDO mass and the bulge 
luminosity at the 99.99\% significance level (Table~\ref{tabstatmdo}). 
For our Galaxy, we plot the radio flux both at 10~mas resolution and at 
$\sim\,$10{\arcmin} (23~pc) resolution; the latter is well matched to the linear
resolution for the other sources. At this matched resolution (which includes 
emission from the whole Sgr~A complex), the Galaxy is no longer an outlier in 
the correlations. 
 
Visually, the relationship between radio power and bulge luminosity appears
to have less scatter than that between radio power and MDO mass, mainly
because of the outliers in the latter: the massive ellipticals at the highest
radio powers and NGC~3115.
For \mmdo$\,\simeq\,10^{8.5-9}$, there is a range of a factor of 10$^5$ in 
P$_{\rm 2cm}$. 
We compare the correlations between the three variables using
the Kendall $\tau$ coefficient. Numerically, 
$P_{\rm concordance}/P_{\rm discordance}$ = (1+$\tau$)/(1$-\tau$), where
$P$ is the probability. 
That is, $\tau=1$ implies perfect concordance (i.e. perfect proportionality), 
$\tau=-1$ implies complete discordance (i.e. perfect inverse proportionality),
and $\tau=0$ implies an absence of correlation.
If we consider all the galaxies in Fig.~\ref{mdorad},
the Kendall $\tau$ coefficient indicates that
the closest correlation is between black hole mass and bulge luminosity
(Kendall's $\tau$= 0.53), while the radio power is more or less equally correlated
with the black hole mass ($\tau$= 0.32) and bulge luminosity ($\tau$= 0.29).
The significance of all three correlations is 99.9\% (Table~\ref{tabstatmdo}).
In order to determine the primary correlation between the three quantities  
(radio luminosity, black hole mass and bulge luminosity), we use a partial 
correlation analysis based on the Kendall $\tau$ coefficient \citep{akrsie96}. 
This method determines the partial correlation between the independent and 
dependent variables after taking away the influence of any correlation with a 
third (test) variable.
We get similar results for both cases - partially correlating \mmdo\ and radio 
power with L(bulge) as the test variable (partial $\tau$= 0.20, significance 
95\%), and partially correlating L(bulge) and radio power with \mmdo\ as the 
test variable (partial $\tau$= 0.16, significance 95\%).
The presence of a 95\% significance for the partial correlations indicates 
that the radio power is dependent on \textit{both} the MDO mass and the 
bulge luminosity. This, however, is only a $\sim\,$2$\,\sigma$ result.
Thus, at this point, we cannot determine whether the radio power is primarily
related to the black hole mass or the bulge luminosity or to both equally.
A more conclusive analysis must await a better determination of the MDO 
versus $\sigma_c$ relationship in lower mass ellipticals and in 
non-ellipticals.

Among the 37 ellipticals the radio power is correlated with the bulge 
luminosity at the 98.5\%-99.5\% significance level and with the MDO mass at 
the 98.5\%-99.8\% level (Table~\ref{tabstatmdo}). 
Linear regression analyses by the Buckley James method in Asurv yields: \newline
log P$_{\rm 2cm}^{\rm elliptical}$ = 2.33($\pm$0.67) log L$_{\rm B}$(bulge) 
                                     $-$ 4.35 \newline
log P$_{\rm 2cm}^{\rm elliptical}$ = 2.11($\pm$0.63) log \mmdo\ + 2.33 \newline
Here, and in the equations below, P$_{\rm 2cm}$ is measured in Watt Hz$^{-1}$,
L$_{\rm B}$ in solar luminosities, and \mmdo\ in solar masses.

For the 48 non-elliptical galaxies which are \textit{detected} in the radio 
at resolutions better than 150~mas, the radio power is correlated with
both the MDO mass and the bulge luminosity at the 99.99\% with the
following linear regression fits  \newline
log P$_{\rm 2cm}^{\rm detected~S0/later}$ = 1.92($\pm$0.27) 
                    log L$_{\rm B}$(bulge) + 1.14 \newline
log P$_{\rm 2cm}^{\rm detected~S0/later}$ = 1.15($\pm$0.21) 
                    log \mmdo\ + 11.35 \newline
The presence of a few upper limits to the MDO mass and/or bulge luminosity 
among the non-elliptical galaxies which are not detected in the radio 
complicates the statistical tests. When both variables have upper limits
to some data points, only the Schmitt binned regression can be used for linear 
fitting within the Asurv package. 
The results in this case are quite dependent on the number of 
bins chosen, e.g. for all galaxies with both MDO estimates and radio powers, 
P$_{\rm 2cm}\,\propto\,$M$_{\rm MDO}^{0.8}$ 
when 5 bins are used, and M$_{\rm MDO}^{0.6}$ when 6 bins are used. 
We therefore delete the four nuclei with MDO mass upper limits and the three 
nuclei with upper limits to their bulge luminosities from the sample.
Having done this, the Buckley James linear regression method gives the 
following for all (i.e. whether or not they are detected in the radio) 
non-elliptical galaxies: \newline
log P$_{\rm 2cm}^{\rm all~S0/later}$ = 1.56($\pm$0.30) log L$_{\rm B}$(bulge) 
                                   + 2.85 \newline
log P$_{\rm 2cm}^{\rm all~S0/later}$ = 1.07($\pm$0.21) log \mmdo\ + 10.52 \newline

The standard deviations of the two relationships are similar for the 
ellipticals and non-ellipticals. Even though the linear regression gives a 
smaller slope for the non-ellipticals than the ellipticals in both 
relationships, this result is not certain given the large number of radio 
upper limits at the lower MDO masses and bulge luminosities. These 
data points could steepen the relationship for non-ellipticals and bring it 
closer to that for ellipticals. In fact, the current data show no contradiction
to all of the galaxies being described by the same relationship. If we do 
this -- i.e. fit ellipticals and non-ellipticals together -- we get: \newline
log P$_{\rm 2cm}^{\rm all}$ = 1.89($\pm$0.21) log L$_{\rm B}$(bulge) 
                                     $-$  0.17 \newline
log P$_{\rm 2cm}^{\rm all}$ = 1.31($\pm$0.16) log \mmdo\ + 8.77  \newline
if all galaxies are considered; and  \newline
log P$_{\rm 2cm}^{\rm radio~core}$ = 1.72($\pm$0.20) log L$_{\rm B}$(bulge) 
                                     $+$  3.15 \newline
log P$_{\rm 2cm}^{\rm radio~core}$ = 1.14($\pm$0.16) log \mmdo\ + 11.49  \newline
if only those galaxies with radio cores detected at resolution $\leq$~150~mas 
are considered.

\section{Discussion}
\label{secdiscussion}

For all 16 LLAGNs at \d19 observed at mas resolution 
(Sects.~\ref{resvlba}, \ref{secdefinite}) four factors point strongly 
to a non-thermal origin of their mas-scale radio cores: 
(1)~the brightness temperatures at 6~cm are too high to be explained by 
    thermal processes; 
(2)~the morphologies of the five LLAGNs with extended mas-scale radio 
    structure are suggestive of pc-scale ``jets'';
(3)~the spectral shapes of the radio cores support a non-thermal origin for 
    the radio emission \citep{naget01,naget02}; and
(4)~significant flux variability is observed and is much more likely in 
    non-thermal than thermal sources.
Thus, the mas-scale radio emission is almost certainly related to accretion 
onto a massive black hole.
The high-brightness temperatures of the cores and the presence of either
a single component, or jet structures about a single bright component, make 
it likely that the radio cores we detect are within $\simeq\,1\,$pc of the 
accreting black hole. 
If we are indeed looking directly at the accreting region or the base of
the jet in the radio, then these nuclei are unique among nearby galaxies and 
merit further study in the radio.

The radio results imply that a large fraction (perhaps all) of LLAGNs 
have accreting massive black holes. 
If we consider only the mas detections, then at least 25\%~$\pm$~6\% of
LINERs and low-luminosity Seyferts have accreting black holes.
VLA-detected compact radio cores with flux $<\,$2.7~mJy were not investigated 
with the VLBA; in other respects these cores are similar to those with 
detected mas scale structure. Thus it is likely that \textit{all} LLAGNs with 
VLA-detected compact radio cores (38\%~$\pm$~8\% of LINERs and low-luminosity 
Seyferts) have accreting black holes.
The scalings between radio power, emission-line luminosity, and galaxy
luminosity provide evidence that the radio non-detections are simply
lower power versions of the radio detections.
In fact we find no reason to disbelieve that \textit{all} 
LLAGNs have an accreting black hole.

One of the most important results in this paper is that compact radio cores 
are found almost exclusively in massive ellipticals and type~1 LLAGNs. 
For massive ellipticals, the high bulge luminosity and black hole mass appear
to be key factors related to the production of a radio core, in light of the 
scalings seen between radio power and these parameters. Among non-ellipticals,
the preferential detection of type~1 LLAGNs may result from the limited 
sensitivity of optical and radio observations, which detect broad \ha\ and 
radio cores in only the more luminous LLAGNs.
For example, it may be that type 1 LLAGNs are in an outburst phase, during 
which accretion power contributes significantly to the total nuclear power 
output, and during which they temporarily host both broad \ha\ emission and 
a compact radio core. This scenario is supported by the findings that 
the observed 2--10~keV luminosity in type~2 LLAGNs is insufficient to
power the optical emission-lines unless the X-rays are heavily absorbed
\citep{teret00}, and that the nuclear UV emission in some type~2 LINERs, 
the far-UV component of which is responsible for powering the emission lines, 
is dominated by massive-star clusters \citep{maoet98}. 
As another alternative, one can invoke the unified scheme \citep{ant93} and 
posit that all LLAGNs have accreting black holes and either
(a)~the radio emission in type~1 LLAGNs is beamed (weakly relativistic jets 
   [$\gamma\,\sim\,2$] can give boost factors of up to $\sim\,$5) and/or
(b)~the 2~cm radio emission in type 2 LLAGNs is free-free absorbed by a
    `torus' i.e. $\tau_{\rm 15GHz}\,\geq$~1.

For the distance-limited (\d19) sample of LLAGNs, we found 
(Sect.~\ref{secinterplay}) that the emission-line luminosity increases with 
the galaxy disk luminosity, while the radio power is primarily correlated 
with the galaxy bulge luminosity. Thus
doubling the bulge luminosity almost doubles the 
P$_{\rm rad}$/L$_{\rm emission-line}$ ratio (Sect.~\ref{secgalbulge},
Figs.~\ref{radvso1}, \ref{radvshan2}). 
The lower bulge luminosity of late-type galaxies implies both a smaller radio 
power \textit{and} a smaller P$_{\rm rad}$/L$_{\rm emission-line}$ ratio than
found for ellipticals.
Thus with increasing total galaxy luminosity, the 
P$_{\rm rad}$/L$_{\rm emission-line}$ ratio of elliptical and non-elliptical 
galaxy nuclei will become increasingly separated and can create a trend not 
unlike the `radio-loud' and `radio-quiet' sequences seen in more powerful 
AGNs \citep[e.g.][]{xuet99}.

\begin{figure*}
\resizebox{14.5cm}{!}{        
\includegraphics{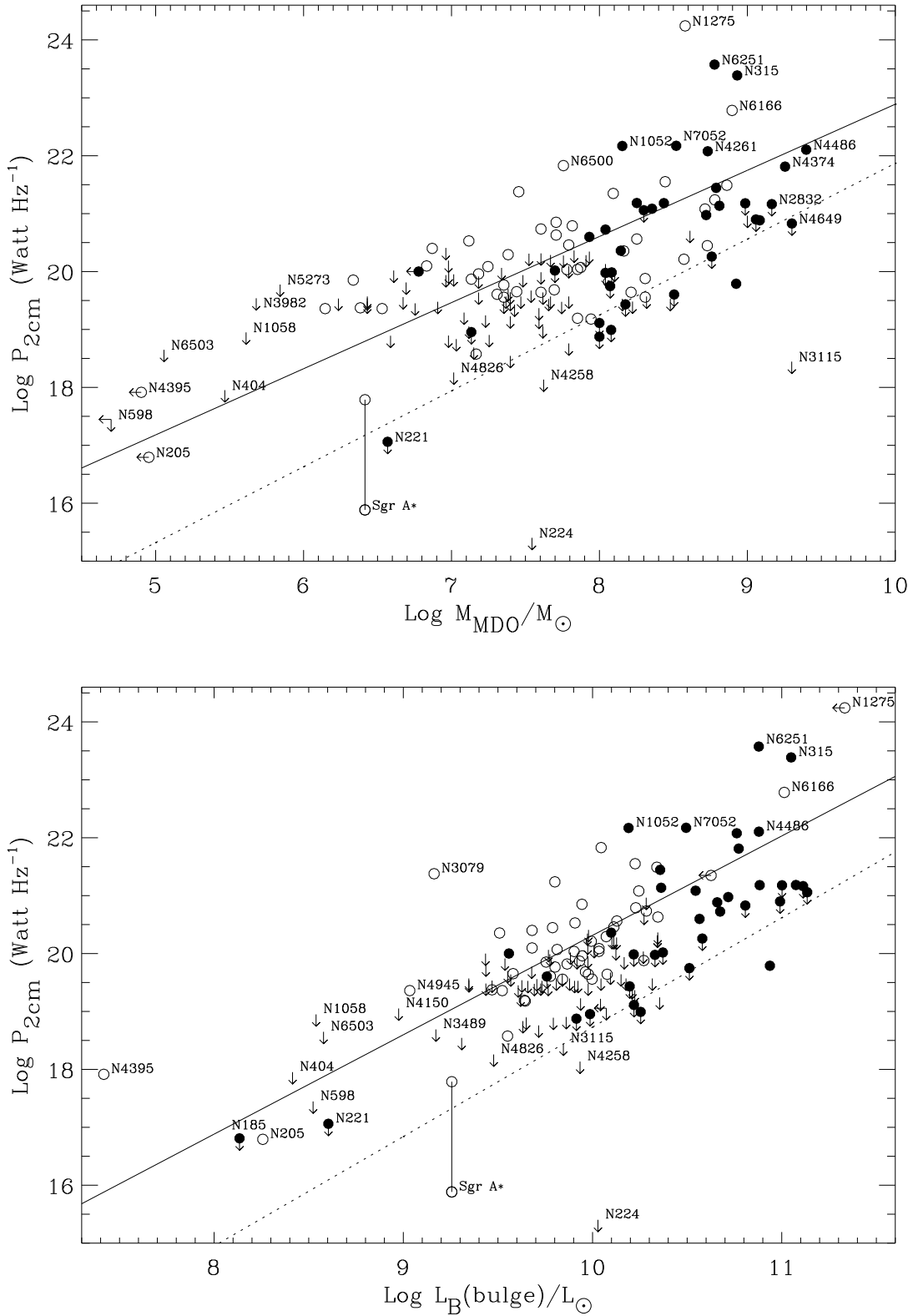}}
\caption{
 \textbf{Upper Panel:}
 The dependence of the core radio power on the mass of the supermassive 
 black hole for nearby galaxies. Only galaxies with all three of 
 radio power, MDO estimate and L$_{\rm B}$(bulge) are plotted.
 Radio flux measurements with resolution $>\,150\,$mas are plotted as
 upper limits.
 Ellipticals are plotted as filled circles and non-elliptical radio
 detections as open circles. For clarity the symbols for non-elliptical nuclei 
 with radio upper limits are not plotted (only the downward pointing arrows are
 plotted for these). Galaxy names are provided for data points in the less crowded 
 areas of the plots. NGC~185, with an estimated MDO mass of 
 6.8$\times10^3\,$\Msun, lies off the left of the plot.
 The two values for \sgra\ correspond to radio powers within 10~mas and
 within $\sim\,$10{\arcmin}.
 In each panel, the solid and dashed lines show the linear fits to all (66) nuclei 
 with radio cores detected at $\leq$~150~mas resolution, and all (154) nuclei 
 plotted, respectively.
 The linear fits take account of the upper limits in both variables.
 \textbf{Lower Panel:} same as upper panel, but for bulge luminosity in the B band
 instead of black hole mass.}
\label{mdorad}
\end{figure*}

In our larger, `MDO' sample (Sect.~\ref{secmdosample}), the radio power is 
correlated with both the estimated MDO mass 
(P$_{\rm 2cm}^{\rm all}\,\propto$  M$_{\rm MDO}^{1.3}$)
and the luminosity of the galaxy bulge 
(P$_{\rm 2cm}^{\rm all}\,\propto$ L$_{\rm B}^{1.9}$(bulge))
over five or more orders of magnitude (Fig.~\ref{mdorad}).
At this point it is not clear which of the two is the primary correlation. 
Partial correlation analysis suggests (at the $\sim\,2\sigma$ level) that 
\textit{both} correlations are physically significant. If so, the latter 
correlation could be due to the influence of the bulge mass on the accretion 
rate. More complete results must await a better characterization of the MDO 
versus $\sigma_c$ relationship in low mass ellipticals and in non-ellipticals.

One model which may account for the correlation between radio power and black 
hole mass posits that the radio emission is thermal synchrotron
radiation from the accretion inflow in an ADAF or CDAF type model. The 
approximate proportionality between these two 
quantities (Sect.~\ref{secmdo}) is consistent with the
L$_{\rm 2\,cm}\,\sim\,3\times10^{36}\,M_{7}^{8/5}\,\dot{m}_{-3}^{6/5}$ ergs~s$^{-1}$
relationship expected in the ADAF model of \citet{yibou99} for values of 
$\dot{m}$ $\sim$ 10$^{-2}$ to $\sim\,$10$^{-5}$ of the Eddington accretion rate.
Here $\dot{m}_{-3}\,= \dot{m}/10^{-3}$, $\dot{m}$ is the mass accretion rate in
units of the Eddington rate, and $M_{7}$ is \mmdo/$10^7$.
A strong argument against all the radio emission coming from an ADAF 
or CDAF is that many of the galaxies considered here show morphological evidence
for pc-scale jets, have core radio spectra which are flatter than what is expected 
in simple ADAF or CDAF models \citep{naget01} and/or turn over in the 10-30\,GHz range 
(Di Matteo et al. 2001; Nagar et al., in preparation, see Nagar et al. 2002).
If a significant (perhaps dominant) component of the radio emission is from 
jets, then the data in Fig.~\ref{mdorad} imply accretion rates much lower than 
$\sim$\,10$^{-2}-10^{-5}$ of the Eddington rate within an ADAF or CDAF type model.

In the context of a jet model, the correlations between core radio power
and both black hole mass and galaxy bulge luminosity imply either
(a)~the strength of the pc-scale radio jet scales with the mass of the black 
    hole, and/or 
(b)~the radio emission from the pc-scale jet is influenced by the 
    potential of the galaxy bulge. 
In the case of (a), if the radio power is linearly dependent on the black hole 
mass, as is consistent with Fig.~\ref{mdorad}, then 
\citep[following the arguments of][]{ho02} one can estimate the radio 
luminosity empirically from the bolometric luminosity (L$_{\rm Bol}$) of the AGN. 
This method implies that the nuclei with detected radio cores in 
Fig.~\ref{mdorad} have L$_{\rm Bol}$/L$_{\rm Edd} \sim\,10^{-2}$ to 10$^{-4}$,
with the linear fit to the radio detections (solid line in upper panel of 
Fig.~\ref{mdorad}) lying approximately at
L$_{\rm Bol}$/L$_{\rm Edd} \sim\,10^{-3}$. 
The maximal jet model of \citet{falbie96} predicts that 
P$_{\rm 2cm}^{\rm core}\,\propto \dot{m}^{1.4}$: to fit the 
slope of the data in Fig.~\ref{mdorad} this requires $\dot{m}$ to
scale roughly linearly with \mmdo, equivalent to requiring the LLAGNs to
accrete at a roughly fixed fraction of the Eddington rate. 
If $\dot{m}$ does scale with \mmdo\, this model can reproduce the 
observed radio luminosities of the nuclei in Fig.~\ref{mdorad} even for
L$_{\rm Bol}$/L$_{\rm Edd} \sim\,10^{-3}$
(comparing Fig.~\ref{mdorad} with Fig.~1 of Falcke \& Biermann and
 assuming their L$_{\rm disk}\,\sim$ L$_{\rm Bol}$).

In the context of (b) above, the relationship between radio power and black 
hole mass may come about indirectly as outlined below.
The core radio power, at arcsec-scale resolution, is strongly
correlated with the bulge magnitude over 6 orders of radio power for FR~I 
and FR~II radio galaxies, and Seyfert galaxies \citep{nelwhi96}.
For several reasons, powerful radio galaxies and luminous Seyfert galaxies are 
not posited to be powered by low-radiative-luminosity models, e.g. their 
derived accretion rates can be $\dot{m}\,\gtrsim\,$10$^{-1.6}$, which is the 
currently believed upper limit for an ADAFs existence \citep{naret98}, and they 
show rapid X-ray variability \citep[e.g.][]{ptaet98}.  
Thus, it is possible that some other factor is responsible for the scaling of
the core radio power with the host galaxy bulge luminosity. This factor could
be the mass loss rate of stars in the bulge, which should be related to both the
black hole accretion rate and the pressure of the interstellar medium in the
vicinity of the nucleus. A higher thermal pressure around the radio source would
tend to confine it and enhance the radio power \citep[cf.][]{nelwhi96}.
Alternatively, the correlation with black hole mass could be the primary one. 
As we have shown, the nearby galaxies we consider in Fig.~\ref{mdorad} 
(lower panel) follow P$_{\rm 2cm}^{\rm core}\,\propto\,$L$^{1.9}_{\rm B}$(bulge).
Also the black hole mass scales with the bulge mass as
M$_{\rm MDO}~= 0.005~$M$_{\rm bulge}$, where
M$_{\rm bulge}~= 2.5 \times 10^7$ (L$_{\rm bulge}/10^9\,$L$_{\odot}$)$^{1.2}$
\citep{ricet98}. Combining these equations gives
P$_{\rm 2cm}^{\rm core}\,\propto\,$M$_{\rm MDO}^{1.6}$.
The slope of this relationship is exactly the same as that of the ADAF model of
\citet{yibou99} and is not too far from the behavior of the data in 
Fig.~\ref{mdorad} (upper panel).

\section{Conclusions}

The completion of our radio survey of the 96 nearest (\d19) LLAGNs
from the Palomar sample of nearby bright galaxies has yielded
the following main results: \newline
a)~All (16 of 16) of the LLAGNs investigated at mas resolution with the VLBA
   have pc-scale cores with brightness temperatures $\gtrsim\,10^8$~K. 
   The five of these nuclei with the highest core flux
   all have pc-scale jets. The luminosity, brightness temperature, morphology, and 
   variability of the radio emission all argue against an origin in star-formation 
   related processes or as thermal emission. Thus, the core radio emission probably
   originates either in an accretion inflow onto a supermassive black hole or
   from jets launched by this black hole-accretion disk system. The latter
   explanation is supported by the morphologies and spectra of the radio cores; \newline
b)~there is no reason to believe that the remaining LLAGNs with compact radio 
   cores (investigated at 150~mas resolution) are different from the 16 LLAGNs 
   investigated at mas resolution.
   Thus, at least half of all LINERs and low-luminosity Seyferts probably contain
   accreting black holes. The incidence for transition nuclei is much lower; 
   \newline
c)~compact radio cores are preferentially found in massive ellipticals and
   in type~1 nuclei, i.e.  nuclei in which broad \ha\ emission is present. 
   For these nuclei, the core radio power is proportional to the broad \ha\
   luminosity.
   The preferential detection of type~1 nuclei could result from: 
   1) observational selection effects, in which broad \ha\ 
   and radio cores have been found only in the more powerful LLAGNs, 
   2) only the type~1 LLAGNs are bona-fide AGNs, or 3) within the unified scheme, 
   type~1 nuclei are beamed and/or type~2 nuclei are free-free absorbed in the 
   radio; \newline
d)~the radio power of the compact core is correlated with both the galaxy 
   luminosity and the luminosity and width of the nuclear emission lines.
   These trends suggest that we have detected only the brighter LLAGNs, i.e. the 
   true incidence of accreting black holes in LLAGNs is higher than indicated by
   our radio detections alone. \newline
e)~The core radio and nuclear emission-line properties of LLAGNs fall close to the 
   low-luminosity extrapolations of more powerful AGNs, providing further support 
   for a common central engine;  \newline
f)~low-luminosity Seyferts and LINERs share many of the same characteristics in 
   the radio. The transition nuclei detected are those which are the closest, in 
   terms of emission-line diagnostic ratios, to Seyferts and LINERs. 
   Thus at least some transition nuclei are really composite 
   Seyfert/LINER + \hii\ region nuclei, with 
   the core radio power dependent on the Seyfert/LINER component; \newline
g)~with increasing bulge luminosity, the radio power increases more rapidly
   than the emission-line luminosity. Also, there is evidence that the disk 
   luminosity of the galaxy is correlated with the nuclear emission-line 
   luminosity (but not the core radio power). Both these factors contribute to a 
   radio loud/quiet division between ellipticals and non-ellipticals at high
   bulge luminosities; \newline
h)~investigation of a sample of $\sim\,$150 nearby bright galaxies, most of them 
   LLAGNs, shows that the nuclear ($\leq$150~mas) radio power is strongly 
   correlated with both the estimated MDO mass and the galaxy bulge luminosity.
   Partial correlation analysis on the two correlations yields the result that 
   each correlation is meaningful even after removing the effect of the other 
   correlation. The nature of these important correlations is discussed.
   Low accretion rates ($\leq\,$10$^{-2}$--10$^{-3}$ of the Eddington rate) are 
   implied in both ADAF- and jet-type models; \newline
i)~about half of all LLAGNs investigated show significant inter-year variability
   at 2~cm and 3.6~cm.  \newline
In short, all evidence points towards the presence of accreting black holes in 
   a large fraction, perhaps all, of LLAGNs. 

\acknowledgements
NN thanks Tim Heckman for valuable suggestions and corrections, and Alison Peck
for help with the VLBA data reduction. We thank Loretta Gregorini for suggestions
which improved the organization of the paper.
This work was partially supported by the Italian Ministry for University and 
Research (MURST) under grant Cofin00-02-36 and the Italian Space Agency (ASI) 
under grant 1/R/27/00.
Part of this paper is drawn from NN's thesis work at the University of 
Maryland, College Park.

\clearpage

\clearpage

\clearpage

\clearpage

\clearpage

\clearpage

\clearpage

\clearpage

\clearpage

\begin{table}
    \dummytable\label{tabll16}
\end{table}
\begin{table}
    \dummytable\label{tabll96}
\end{table}
\begin{table}
    \dummytable\label{tabvlbanew}
\end{table}
\begin{table}
    \dummytable\label{tabstatll96}
\end{table}
\begin{table}
    \dummytable\label{tabstatmdo}
\end{table}

\begin{figure}
\includegraphics[width=7cm,clip]{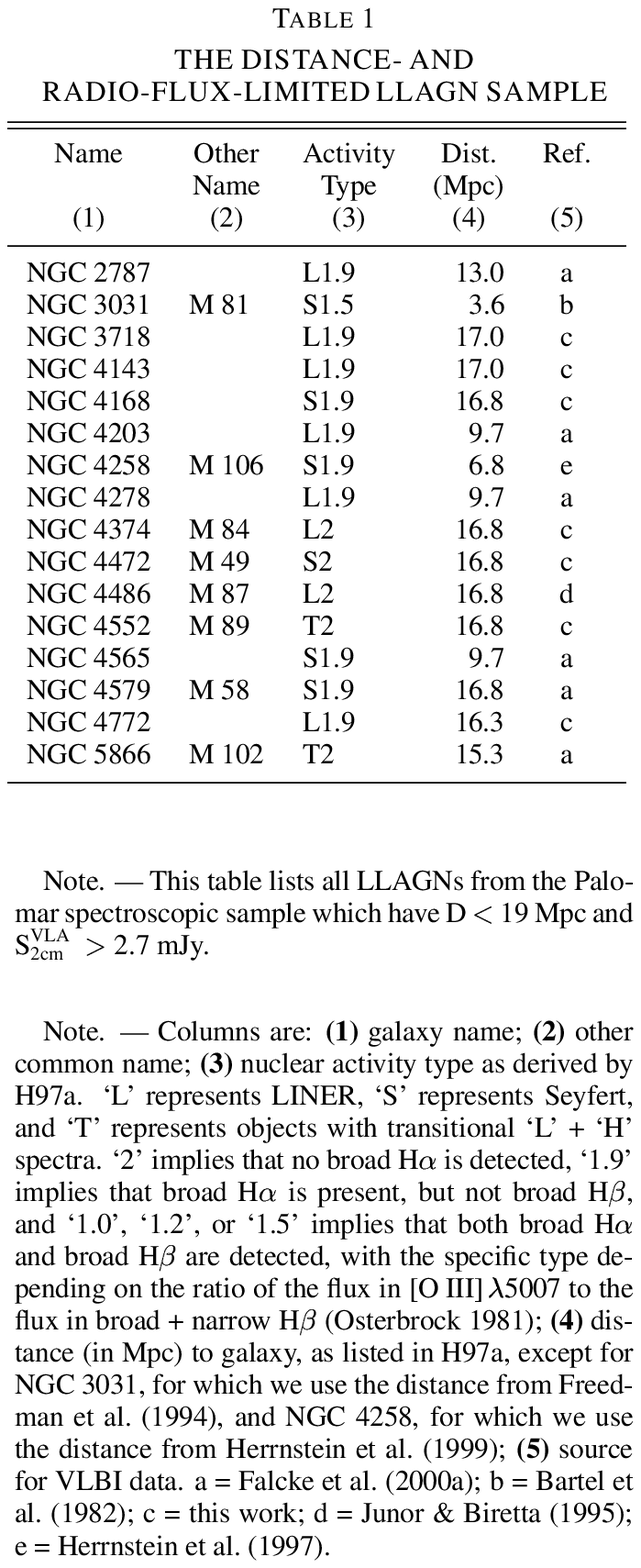}
\end{figure}

\clearpage

\begin{figure*}
\includegraphics[width=\textwidth,clip]{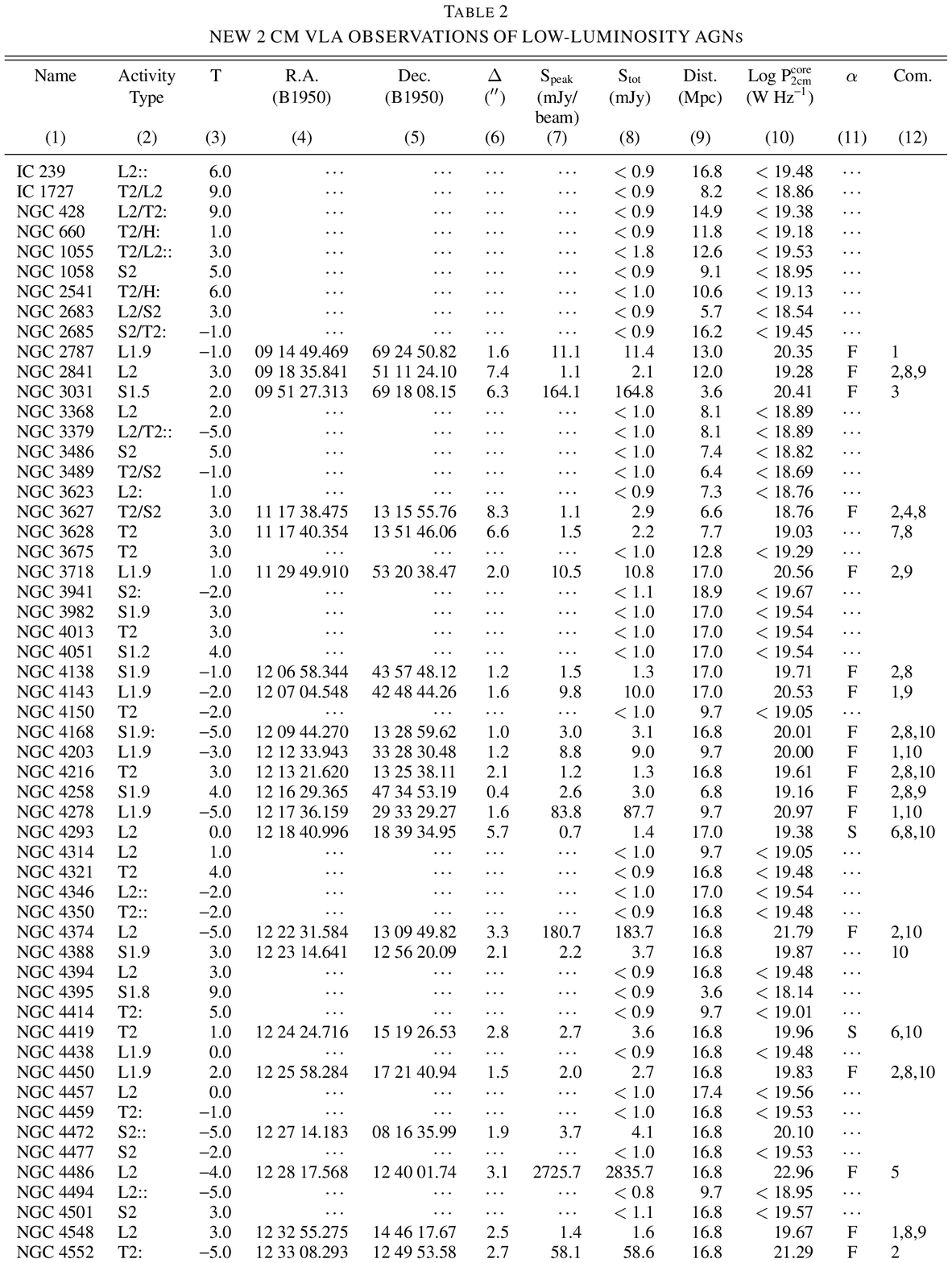} 
\end{figure*}

\clearpage

\begin{figure*}
\includegraphics[width=\textwidth,clip]{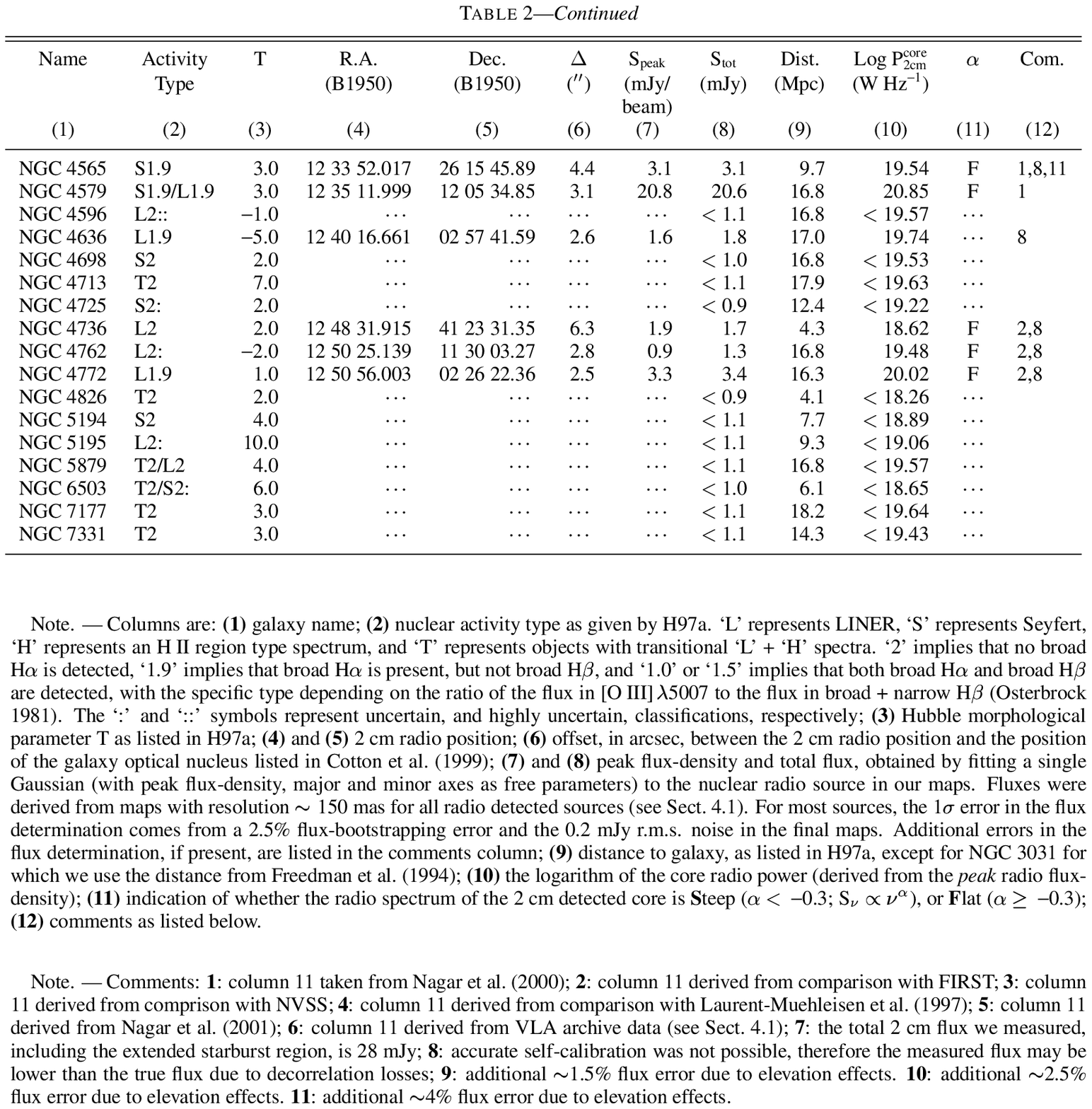} 
\end{figure*}

\clearpage

\begin{figure*}
\includegraphics{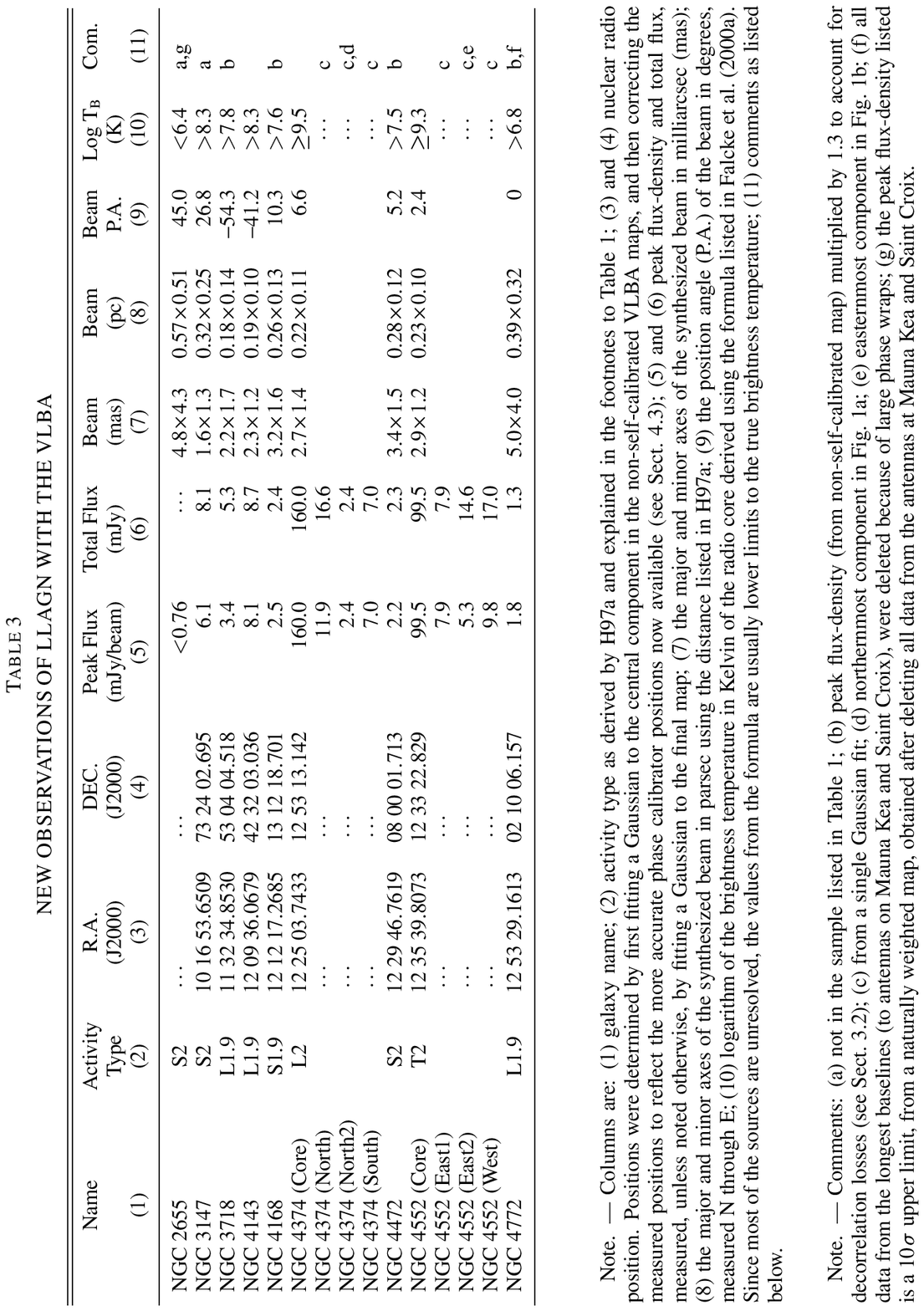}
\end{figure*}

\clearpage

\begin{figure*}
\includegraphics[width=\textwidth,clip]{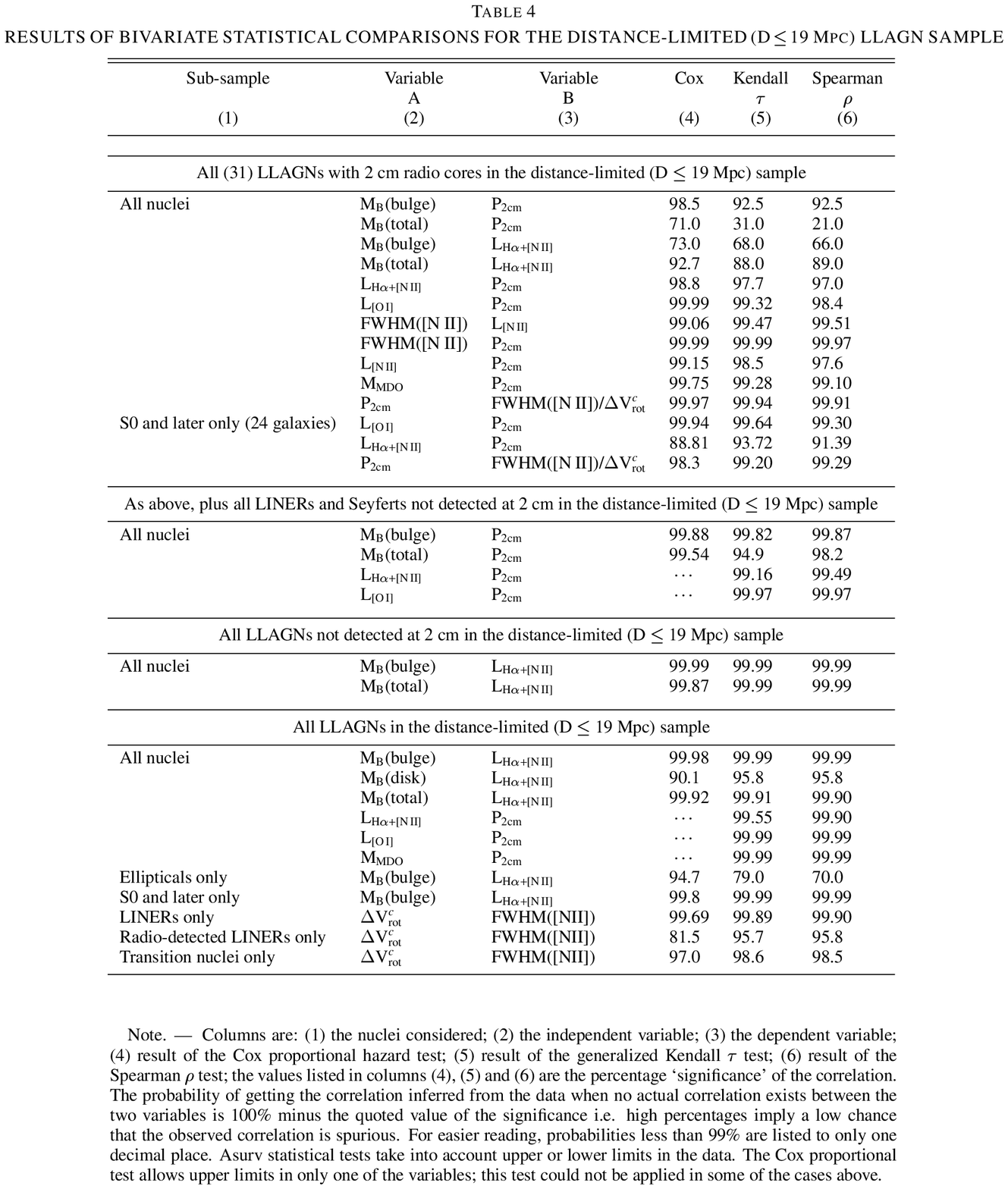} 
\end{figure*}

\clearpage

\begin{figure}
\includegraphics[width=7cm,clip]{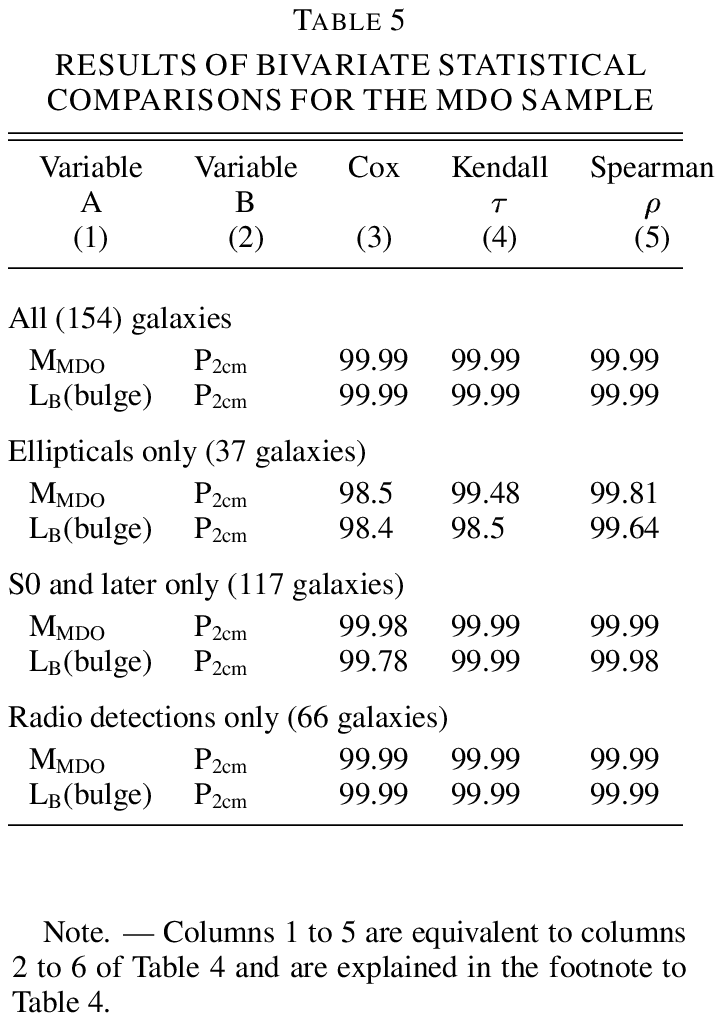}
\end{figure}

\end{document}